\documentclass[preprint,aps,showkeys]{revtex4-1}

\usepackage{amsmath}%
\usepackage{graphicx}%
\usepackage{color}%
\usepackage{dcolumn}%
\usepackage{bm}%
\usepackage{ulem}
\usepackage[uline]{hhtensor}
\usepackage{setspace}
\usepackage{lineno,hyperref}
\begin{document}
%\begin{frontmatter}
\title{Ab initio study of oxygen segregation in silicon grain boundaries: the role of strain and vacancies}
\author{Rita Maji}
\address{Dipartimento di Scienze e Metodi dell'Ingegneria, Universit{\`a} di Modena e Reggio Emilia,Via Amendola 2 Padiglione Morselli, I-42122 Reggio Emilia, Italy}
\author{Eleonora Luppi}
\address{Sorbonne Universit\'e, UPMC Univ Paris 06, UMR 7616, Laboratoire de Chimie Th\'eorique, F-75005 Paris, France. CNRS, UMR 7616, Laboratoire de Chimie Th\'eorique, F-75005 Paris, France.}
\author{Nathalie Capron} 
\address{Sorbonne Universit\'e, CNRS, Laboratoire de Chimie Physique Mati\`ere et Rayonnement, UMR 7614, F-75005 Paris, France.}
\author{Elena Degoli}
\address{Dipartimento di Scienze e Metodi dell'Ingegneria, Universit{\`a} di Modena e Reggio Emilia, Via Amendola 2 Padiglione Morselli, I-42122 Reggio Emilia, Italy, \\ 
Centro Interdipartimentale En$\&$Tech, Via Amendola 2 Padiglione Morselli, I-42122 Reggio Emilia, Italy, \\
Centro S3, Istituto Nanoscienze-Consiglio Nazionale delle Ricerche (CNR-NANO),Via Campi 213/A, 41125 Modena, Italy}
%\date{\today}
%\pacs{?}

\begin{abstract}
\begin{singlespace}
Multi-crystalline silicon is widely used for producing low-cost and high-efficiency solar cells. During crystal growth and device fabrication, silicon solar cells contain grain boundaries (GBs) which are 
preferential segregation sites for atomic impurities such as oxygen atoms. GBs can induce charge carriers recombination significantly reducing carrier lifetimes and therefore they can be detrimental for Si device performance. We studied the correlation between structural, energetic and electronic properties of $\Sigma$3\{111\} Si GB in the presence of vacancies, strain and multiple O segregation. The study of the structural and energetic properties of GBs in the presence of strain and vacancies gives an accurate description of the complex mechanisms that control the segregation of oxygen atoms. We analysed tensile and compressive strain and we obtained that local tensile strain around O impurities is very effective for segregation. We also studied the role of multiple O impurities in the presence of Si vacancies finding that the segregation is favorite for those structures which have restored tetrahedral covalent bonds. The presence of vacancies attract atomic impurities in order to restore the electronic stability: the interstitial impurity becomes substitutional. This analysis was the starting point to correlate the change of the electronic properties in $\Sigma$3\{111\} Si GBs with O impurities in the presence of strain and vacancies. For each structure we analysed the density of states and its projection on atoms and states, the band gaps, the segregation energy and their correlation in order to characterise the nature of new energy levels. Actually, knowing the origin of defined electronic states would allow the optimization of materials in order to reduce non radiative electron-hole recombination avoiding charge and energy losses and therefore improving solar cell efficiency.
\end{singlespace}
\end{abstract}

%\begin{keyword}
%\end{keyword}

%\end{frontmatter}

\maketitle
\setstretch{1}
\section{Introduction}\label{sec:introduction}

Multi-crystalline silicon is widely used for producing low-cost and high-efficiency solar cells. \cite{Sisolarcell2009review,Sisolarcell2008review,Sisolarcellthinfilm2004,Sisolarcellbook2019} During crystal growth and device fabrication, Si solar cells contain grain boundaries (GBs) which can be detrimental for the device performance. \cite{GBeffect1997,GBeffect1990,GBeffect2012} GBs can create deep-energy states which induce charge carriers recombination generating a reverse flux of electrons across the junction that separates the electron- and hole-conducting media. The effect is a significant reduction of carrier lifetimes.  \cite{lifetime2005,PhysRevLett.115.235502,SiyuJPCL2019,Tringe2000} Avoiding the recombination is extremely important for an optimisation of solar cells efficiency. This can be achieved by tuning the energy levels in the materials.\\
A novel view about Si GBs was given by Raghunathan et al. \cite{ragjohgros-nl14} whom predicted that GBs could instead be beneficial to solar energy conversion. They proposed to exploit the electronic properties of GBs in order to design novel and efficient solar cells. However, electronic properties of GBs can be very difficult to tune. In fact, GB structures are very variegated and can easily form vacancies  with deep defect electronic states, accelerating charge dissipation. \cite{Feng2009} Besides, GBs are also preferential segregation sites for various impurity species which can significantly influence the electrical properties. \cite{ZHAO2017599,PhysRevB.91.035309,ohnoapl2013,KasinnoJAP13,pip.2614} Segregation activity can create recombination centers and have a substantial detrimental impact on the conversion efficiency of solar cells. \cite{YuAndreyJAP2015,Peaker2012}

Oxygen atoms are inevitably introduced during solar-cell crystal growth which usually form precipitate recombination centers with different morphology. \cite{Chen2011,OhnoAPL15} Complex mechanisms control the segregation of oxygen atoms at Si GBs which is influenced by the size and the orientation of the grains and also by the presence of vacancies and strain in the GBs.\cite{KasinnoJAP13,YuAndreyJAP2015,OhnoAPL2017,seagerAnnReVMatSci85,OhnoAPL15,ShiJAP2010,PhysRevLett.121.015702,1.1578699,1.98331} Therefore, the possibility or not to engineer GBs for photovoltaics applications, and more generally to improve solar cell efficiency, depends also on their capacity to segregate different types of impurities such as oxygen atoms. 

Understanding the correlation between structural and electronic properties of GBs in the presence of vacancies and interacting atomic impurities is crucial to improve solar cells technology.

In this work we studied from first-principles how the electronic properties of  $\Sigma$3\{111\} Si GB, in the presence of strain and vacancies, are modified by multiple oxygen segregation. We have chosen the $\Sigma$3\{111\} GB because it is the most likely to form in Si solar cell. \cite{Chen2007,KasinnoJAP13,Ohno2017} This is also the most stable GB since there are no dangling bonds and little bond distortion. \cite{SARAU20112264} Its electrical activity and gettering ability is negligible. \cite{ohnoapl2013,SARAU20112264} However, the behaviour of $\Sigma$3\{111\} Si GB drastically changes in the presence of strain and vacancies. \cite{SARAU20112264,OhnoAPL2017} In fact, these dislocations cause lattice distortion around sites which can act as gettering center for impurity. 

Starting from the unstrained $\Sigma$3\{111\}  Si GB, we considered tensile and compressive strain differently applied to the GB. Then, multiple oxygen atoms were introduced in the strained GBs (SGB) in many different possible configurations. For each configuration and type of strain, we calculated the oxygen segregation energy revealing the factors influencing the segregation. The same methodology was then implemented to investigate the oxygen segregation in $\Sigma$3\{111\} Si GB with a Si vacancy (VGB). Starting from $\Sigma$3\{111\} GB with a Si vacancy, we considered multiple oxygen atoms in different structural configurations within the GB, and also in this case we studied the mechanisms that regulate the oxygen segregation. 

The electronic properties for each structure have been investigated by analysing the density of states (DOS) and the band gaps versus the segregation energy. We characterised the origin of the new energy levels due to the presence of GBs, strain, vacancies and/or oxygen atoms. Actually, knowing the origin of the electronic states permits to find strategies to reduce non radiative electron-hole recombination avoiding charge and energy losses and therefore improving solar cell efficiency.

\section{Methodology and grain boundary structure}
\label{sec:methogb}

The $\Sigma$3\{111\} Si GB consists of two grains of Si forming an interface along the crystallographic plane \{111\} (coincidence site lattice). The two Si grains are misoriented with respect to one another by an angle $\Omega = 60^{\circ}$.  In Fig.~(1) we show the $\Sigma$3\{111\} orthorhombic supercell ($a$ $\ne$ $b$ $\ne$ $c$ and $\alpha$ = $\beta$ = $\gamma$= $90^{\circ}$) composed of 96 Si atoms generated with {\it GB Studio program}. \cite{GBStudio2006}. The lattice parameters are $a$=13.30~\AA, $b$=7.68~\AA~and $c$=18.81~\AA. It is possible to observe the presence of two GBs, one in the middle of the cell and the other at the edges of the cell due to the periodic boundary conditions. 

Then, we inserted O atoms in the $\Sigma$3\{111\} Si GB which was additionally strained (see section: \ref{OinterstitalsStrain}) and in which we also created a vacancy (see section: \ref{OinterstitalsVacancy}). For each structure we calculated the impurity and the segregation energy of oxygen atoms. Moreover, in order to compare all the structures, we took Si bulk as a reference. In this case, we used a cubic supercell ($a$ $=$ $b$ $=$ $c$ and $\alpha$ = $\beta$ = $\gamma$= $90^{\circ}$) of 64 atoms with $a$=10.86~\AA. This value for $a$ was obtained by the experimental lattice constant 5.431 \AA\,\,for a face-centered cubic unit cell of two Si atoms. \cite{PhysRevB.32.3792}  We also report the calculated Si bulk modulus $B$ = 95.4 GPa and the elastic tensor components $c_{11}$ = 159.8 GPa  and $c_{12}$ = 63.2 GPa. These values are in good agreement with other theoretical calculations and with the experimental values $B$ = 99.2 GPa, $c_{11}$ = 167.5 GPa  and $c_{12}$ = 65.0 GPa. \cite{PhysRevB.32.3792}

\begin{figure}[h!]
\centering
\includegraphics[scale=0.4]{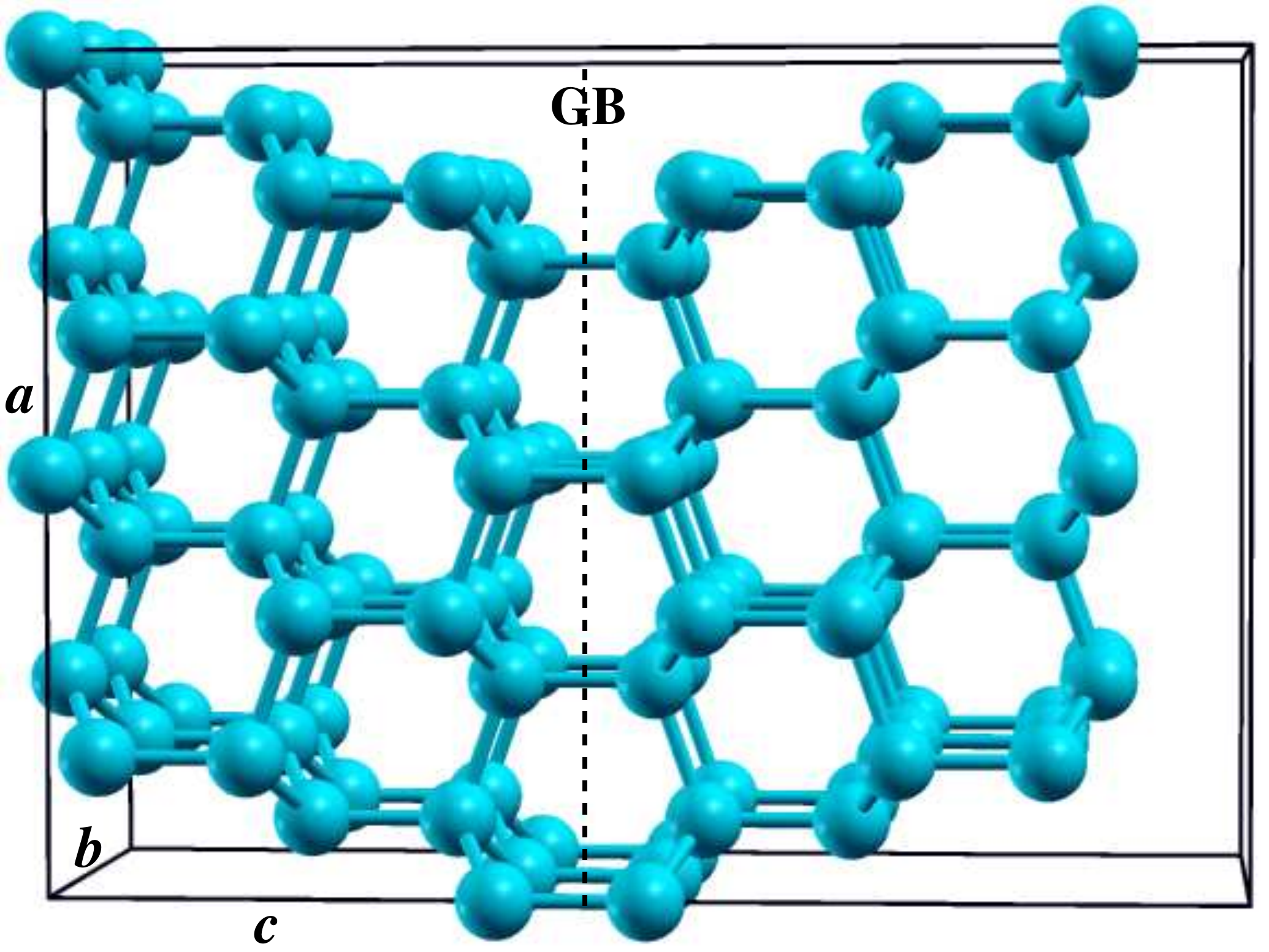}
\caption{$\Sigma$3\{111\} Si GB orthorhombic supercell: $a$, $b$ and $c$ are the lattice parameters. The dotted line shows the GB between two Si grains.}
\label{GB96}
\end{figure}

The calculations were performed using density functional theory (DFT) as implemented in the plane-wave based Vienna Ab initio Simulation Package (VASP). \cite{Hafner, Kresse} We employed the generalised gradient approximation PBE for the exchange-correlation functional \cite{PhysRevLett.77.3865} and projector augmented-wave (PAW) pseudopotentials with a cutoff of 400 eV. K-points sampling within the Monkhorst Pack scheme \cite{Monkhorst} was used for integration of Brillouin-zone together with the linear tetrahedron method including Bl\"ochl corrections. \cite{PhysRevB.49.16223} In particular, we used a k-mesh of 3$\times$3$\times$3 to calculate energy properties of the structures and a k-mesh of 7$\times$7$\times$7 to calculate their density of states (DOS). For the structural optimisation, we used as threshold on the forces the value of 10$^{-2}$ eV/\AA{} per atom.

The choice of the PBE functional was motivated by the need to find a balance between accuracy and computational cost. In the recent years, new GGA functionals for solids have been proposed to improve the description of structural properties. \cite{PhysRevB.80.195109} However, the improvements over PBE are not systematic \cite{PhysRevB.80.195109} and in some cases even less accurate. \cite{PhysRevB.79.155107}

Concerning the electronic properties, it is well known that PBE underestimates the electronic band gaps. For Si bulk we obtain 0.574 eV compared with the experimental value of 1.25 eV. \cite{lupp+08prb}. Hybrids functionals or GW corrections can be used to obtain higher accuracy, but the computational cost is very high for the systems studied here. However, despite the understimation of the absolute values of the gaps, we believe that the trend of the electronic properties is also valid in PBE. For example, for Si bulk the effect of an hybrid functional or GW corrections is to rigidly shift the band structure. \cite{PhysRevB.92.081204,GauJCP2019}

\section{Results and discussion }
\label{sec:resdisc}

\subsection{Oxygen atoms in $\Sigma$3\{111\} Si GB : the role of strain}
\label{OinterstitalsStrain}	
The formation energy of $\Sigma$3\{111\} Si GB is calculated as 
\begin{equation}
E^{\text{f}}_{\text{GB}}  = \frac{E_{\text{GB}} - n_{\text{GB}}e_{\text{B}}} {2 A }, 
\label{EfGB} 
\end{equation}
where $E_{\text{GB}}$ is the energy of the GB, $n_{\text{GB}}$ is the number of atoms (96) in the GB supercell, $e_{\text{B}}$ is the energy per Si atom taken from the Si cubic crystal structure, $A$ is the GB cross-section area and the scaling factor 1/2 denotes the presence of two GBs in the supercell. \cite{PhysRevB.91.035309}  We obtained a very low formation energy $E^{\text{f}}_{\text{GB}} = 0.002$ eV/\AA$^{2}$ ($E^{\text{f}}_{\text{GB}} = 0.05$ J/m$^{2}$) which is the indication of the very regular structure of the $\Sigma$3\{111\} GB, i.e. bond lengths and angles are close to the Si bulk. \cite{ZHAO2017599,PhysRevB.91.035309} 

\begin{figure}[h!]
\centering
\includegraphics[scale=0.4]{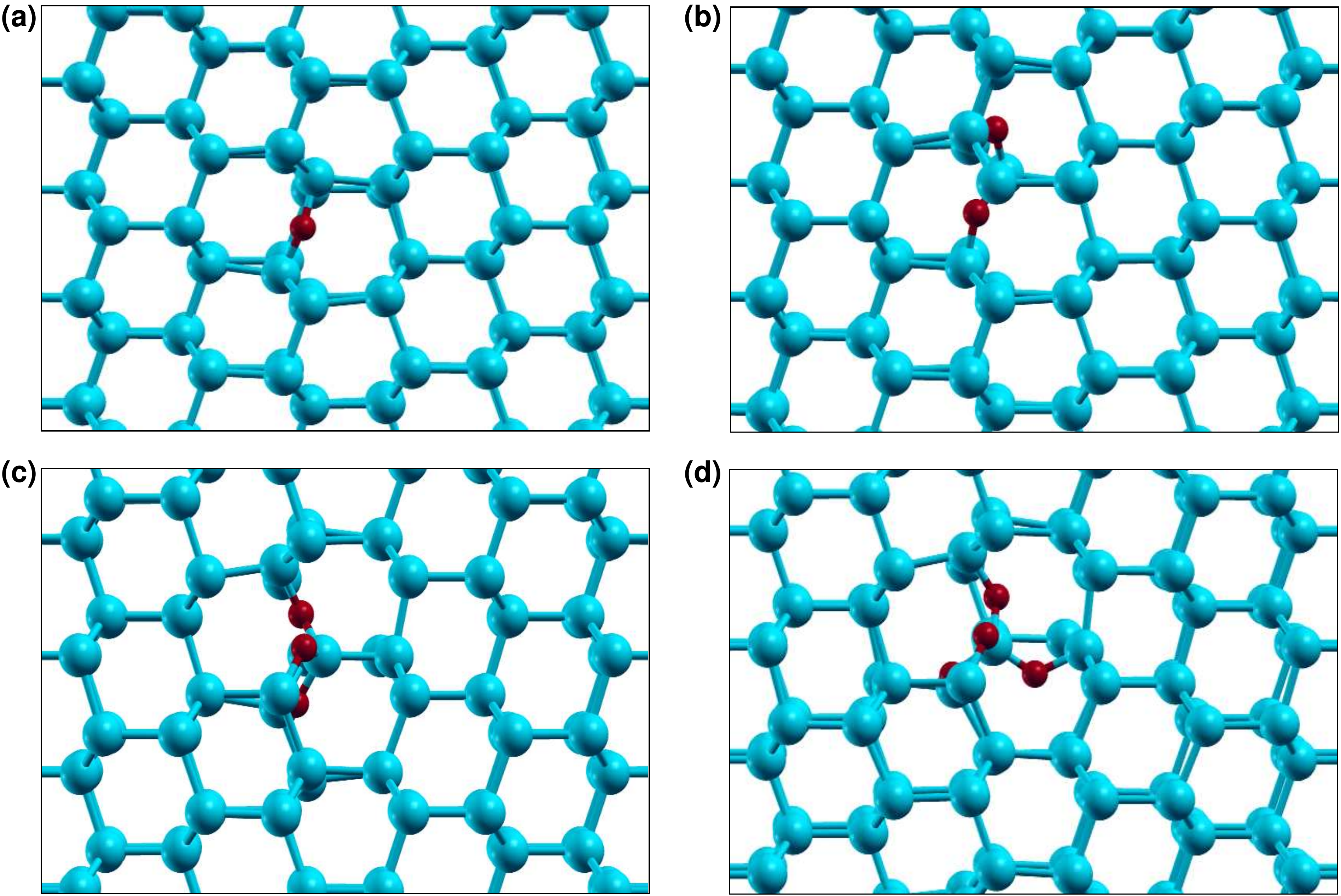}
\caption{$\Sigma$3\{111\} Si GB structures with a number $n$ = 1 (a), 2 (b) , 3 (c)  and 4 (d) of O atoms (red balls) in interstitial configurations. Between all the different structures investigated, here we show those that have the lowest total energy (LE).}
\label{GB_allO}
\end{figure}

We started by inserting O atoms in the $\Sigma$3\{111\} Si GB  in different configurations and we optimised the structures. In the Supp. Mat. we show examples of structures before and after optimisation. We always obtain that the O atoms segregate at bond-centered position between two Si atoms. Therefore, the optimised structures have no coordination defects as all Si atoms restored their tetrahedral coordination (see Fig.~(\ref{GB_allO})). \cite{OhnoAPL2017}  Moreover, the position of the O atoms in the GB can change of sites between the different structures, but the structural parameters (bond lengths and angles) and total energies differ only about 0.01 \%. 

To investigate the interaction between the O atoms and the $\Sigma$3\{111\} Si GB we calculated the segregation energy of the O atoms as 

\begin{equation}
\Delta^{n\text{OGB}}_{n\text{OB}} = E^{n\text{OGB}} - E^{n\text{OB}},
\label{bulkseg} 
\end{equation}

where $E^{n\text{OGB}}$ and $E^{n\text{OB}}$ are the impurity energies of the O atoms respectively in the GB and in the bulk Si which are calculated as

\begin{equation}
E^{n\text{OGB}} = E_{n\text{O+GB}} - E_{\text{GB}}  - n \mu_{\text{O}},
\label{EnOGB} 
\end{equation}

\begin{equation}
E^{n\text{OB}} =  E_{n\text{O+B}} - E_{\text{B}} - n \mu_{\text{O}}.
\label{EnOB} 
\end{equation}

$E_{n\text{O+GB}}$ is the total energy of the GB containing $n$ number of O atoms, $E_{\text{GB}}$ is the total energy of the GB,  $\mu_{\text{O}}$ is the chemical potential of oxygen calculated as the energy per atom of an $O_2$ molecule in vacuum, $E_{n\text{O+B}}$ is the total energy of the Si bulk containing $n$ number of O atoms \cite{PhysRevB.80.144112} and $E_{\text{B}}$ is the total energy of Si bulk. 

\begin{figure}[h!]
\centering
\includegraphics[scale=0.4]{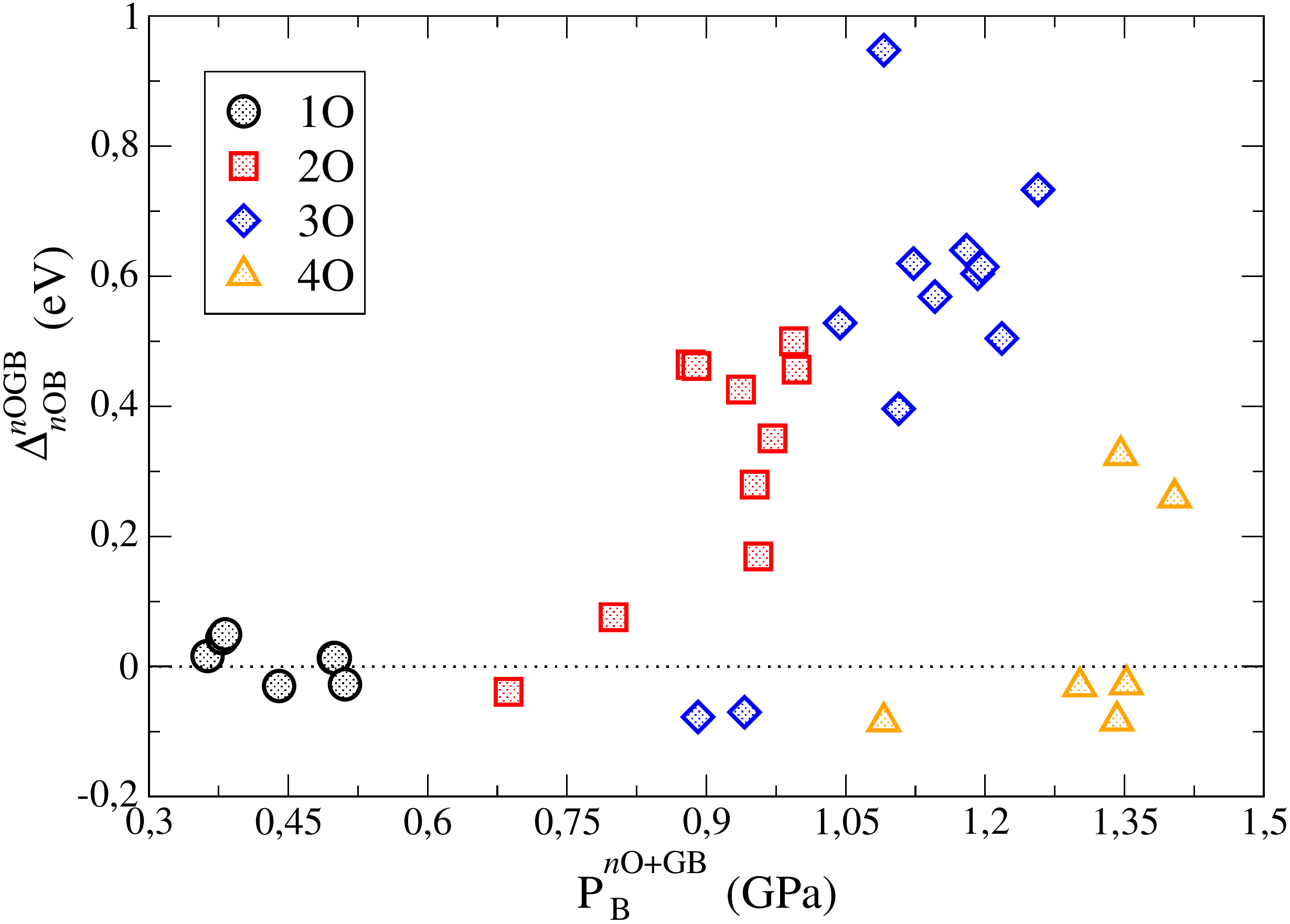}
\caption{O segregation energy $\Delta^{n\text{OGB}}_{n\text{OB}}$ in GB as a function of $P^{n\text{O+GB}}_{\text{B}}$.} The number of O atoms is $n$ = 1, 2, 3 and 4.%The structures with 1 oxygen are in black, with 2 oxygen are in red, with 3 oxygen are in blue and with 4 oxygen are in orange.}
\label{seg_pressure_plot1}
\end{figure}

In Fig.~(\ref{seg_pressure_plot1}) we show the O segregation energies for all the investigated structures which were ordered by their pressure $P^{n\text{O+GB}}_{\text{B}}=P_{n\text{O+GB}} - P_{\text{B}}$, calculated as the difference between the pressure of the GB containing $n$ number of O atoms ($P_{n\text{O+GB}}$) and the pressure of the Si bulk ($P_{\text{B}}=1.973$ GPa). %The value of $P^{n\text{O+GB}}_{\text{B}}$ becomes larger increasing the number of oxygen atoms in the GB because of the increasing stress within the cell.

Most of the structures have positive segregation energies (not larger than 1 eV) while few structures have negative segregation energy (not lower than -0.1 eV). In Tab.~(\ref{SnOGBnOB}) we explicitly report the O segregation energies for $n$ equal to 1, 2, 3 and 4 interstitial O atoms but only for those structures that have the highest and the lowest total energy ($E_{n\text{O+GB}}$). In the Supp. Mat. we reported the values for all the structures presented in Fig.~(\ref{seg_pressure_plot1}).

\begin{figure}[h!]
\centering
\includegraphics[scale=0.5]{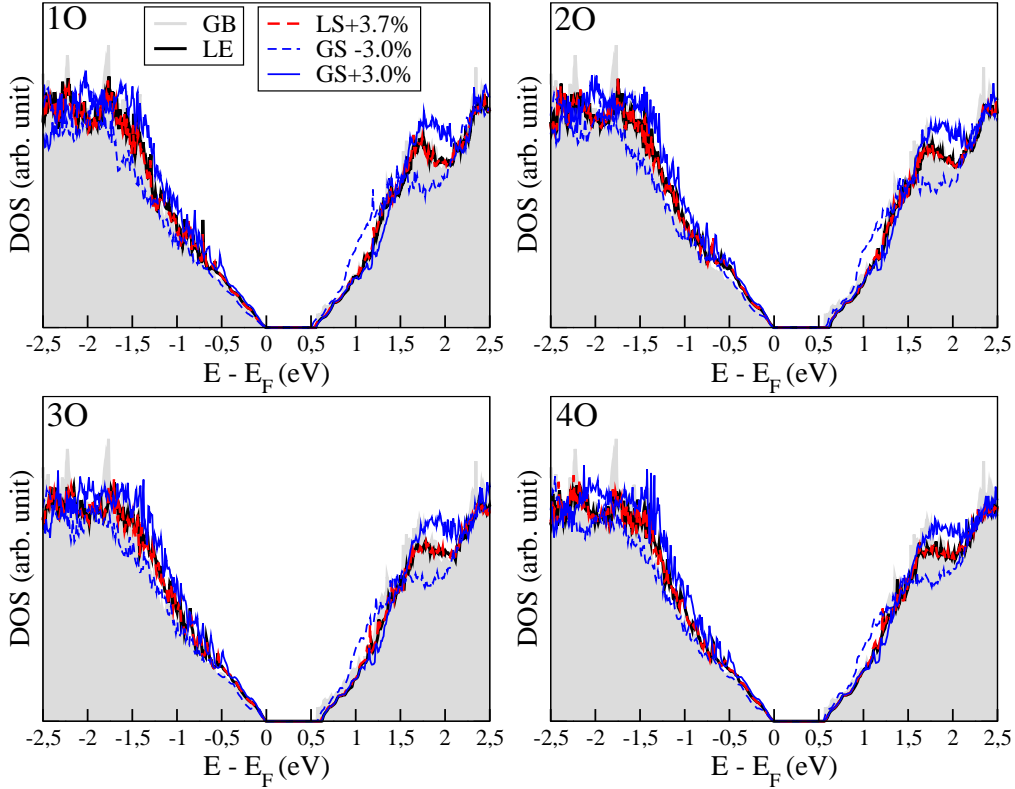}
\caption{Density of states of the GB (grey shadow), LE (black) and SGB with local strain (red) and global strain (blue) structures. The number of O atoms is $n$ = 1, 2, 3 and 4. The percentage of strain is reported in the legend and corresponds to the bond length deformation in the case of local strain and to the lattice parameter deformation in the case of global strain. }
\label{All_DOS_he_le-crop}
\end{figure}

The behaviour we obtained for $\Delta^{n\text{OGB}}_{n\text{OB}}$ shows that oxygen atoms prefer not to segregate at GB as the structures become energetically unfavourable. Therefore, this result confirms that $\Sigma$3\{111\} Si GB has no gettering ability to solute oxygen atoms. \cite{ohnoapl2013,OhnoAPL2016} These structures present very similar energetic properties due to their similar structural properties: this fact impacts also the electronic properties. In Fig.~(\ref{All_DOS_he_le-crop}) we show the DOS for the $\Sigma$3\{111\} Si GB without oxygen atoms (GB) and for $\Sigma$3\{111\} Si GBs with multiple oxygen atoms and lowest total energy (LE). The DOS are almost superimposable. The comparison with the structures with the highest total energy (HE) is reported in the Supp. Mat. showing the same behaviour.

\begin{table}[h!]
\begin{center}
\begin{tabular}{ |c|c|c||c|c||c|c||c|c| } 
\hline
&$P^{\text{1O+GB}}_{\text{B}}$ & $\Delta^{\text{1OGB}}_{\text{1OB}}$ &$P^{\text{2O+GB}}_{\text{B}}$ & $\Delta^{\text{2OGB}}_{\text{2OB}}$& $P^{\text{3O+GB}}_{\text{B}}$ &  $\Delta^{\text{3OGB}}_{\text{3OB}}$&$P^{\text{4O+GB}}_{\text{B}}$ &  $\Delta^{\text{4OGB}}_{\text{4OB}}$ \\ 
\hline
HE  &0.382&  0.049 &0.994&  0.499 &1.091& 0.947 &1.346& 0.325 \\ 
LE  &0.440& -0.030 &0.687&  -0.039 &0.891& -0.078 &1.091& -0.085\\ 
\hline
\end{tabular}
\end{center}
\caption{Oxygen segregation energy $\Delta^{n\text{OVGB}}_{n\text{OGB}}$ (eV) and $P^{n\text{O+VGB}}_{\text{B}}$ (GPa) for $n$ equal to 1,2,3 and 4 oxygen atoms. The values are reported for the LE (lowest total energy) and HE (highest total energy) GB.}
\label{SnOGBnOB}
\end{table}

The energetically unfavourable segregation of oxygen atoms at $\Sigma$3\{111\} Si GB can change in the presence of strain. In fact, a correlation has been found  between the segregation of O atoms and strain. \cite{KasinnoJAP13,OhnoAPL2017} O atoms seem to prefer to segregate under tensile strain. Here we show in detail the ability of strain to segregate oxygen atoms at GB.

\begin{figure}[h!]
\centering
\includegraphics[scale=0.2]{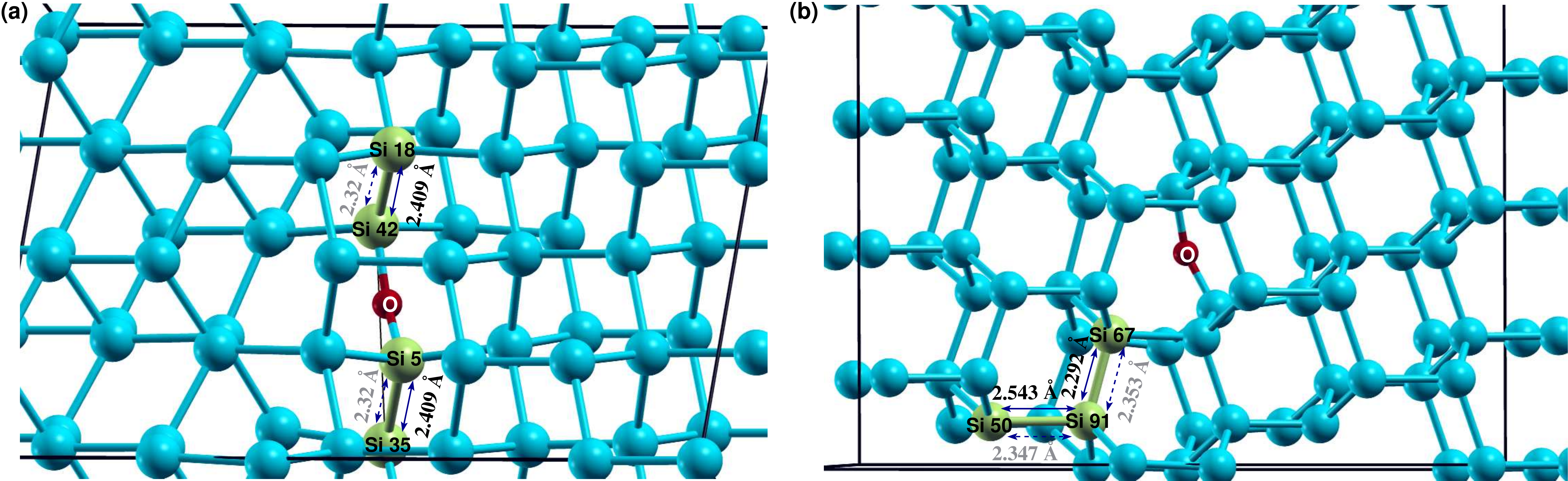}
\caption{$\Sigma$3\{111\} Si GB with one O atom (red ball) under local strain. (a) two bonds close to the O atom are elongated by +3.7\%. (b) two bonds far from the O atom are modified of about -2.6\% and +8.3\%. Modified bond lengths are reported in black (strained) and grey (unstrained).}
\label{strn37}
\end{figure}

\begin{figure}[h!]
\centering
\includegraphics[scale=0.15]{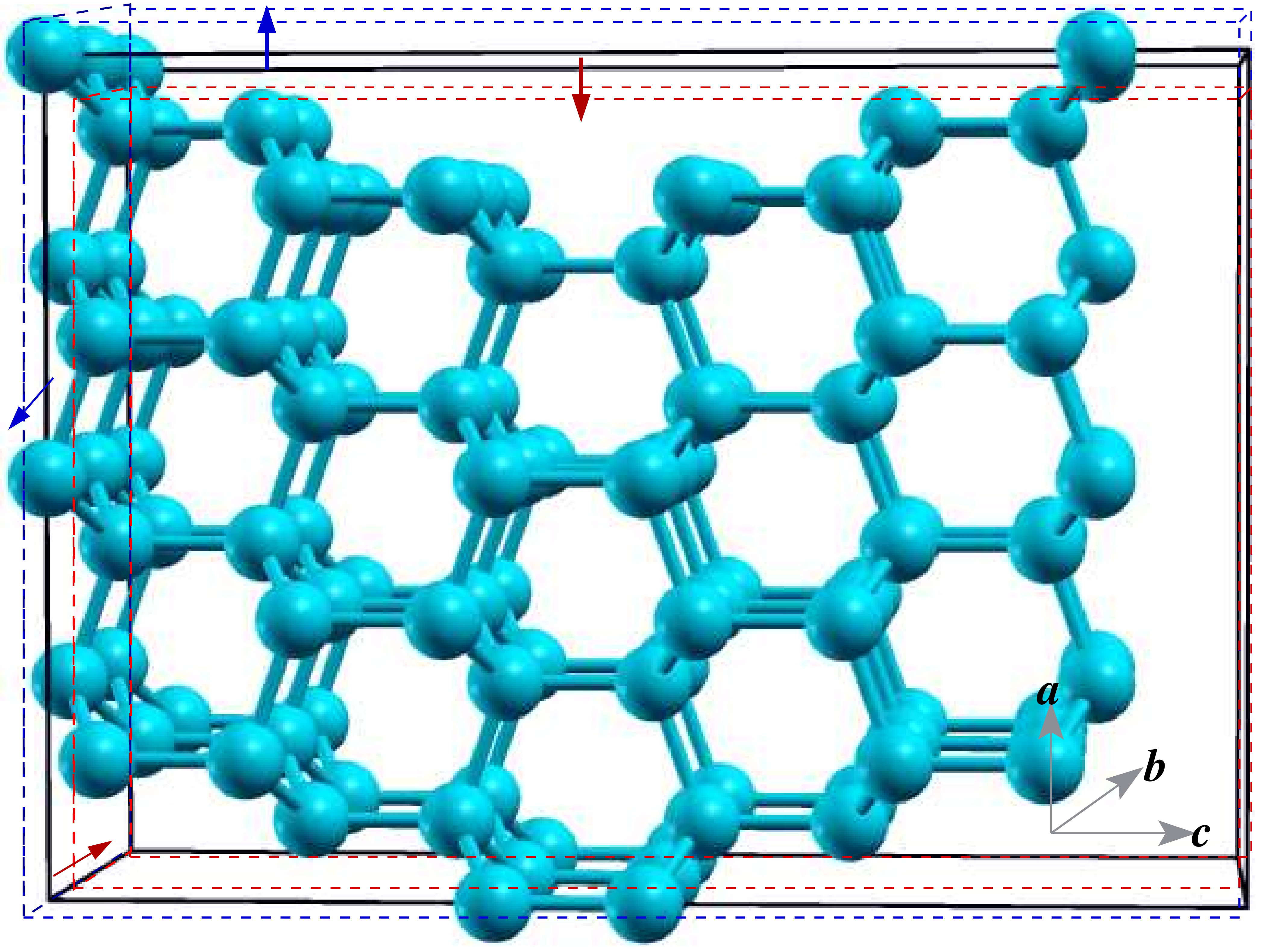}
\caption{$\Sigma$3\{111\} Si GB under global strain. The cell in black-solid line is the unstrained one, while red-dotted line and blue-dotted line represent the compressed and tensile strained cells respectively. The lattice parameter $c$, in the z direction, is left unchanged, while $a$ and $b$, in the xy plane, are modified.}
\label{cell_strain}
\end{figure}

Starting from the unstrained LE $\Sigma$3\{111\} Si GBs previously obtained in the presence of O atoms ($n$ = 1, 2, 3 and 4), we generated two new groups of structures where the strain is simulated in this way:

\begin{itemize}
\item[1)] Local Strain (LS) \cite{PhysRevB.92.075204}: we modified some bond lengths in the LE $\Sigma$3\{111\} Si GB creating locally tensile/compressive strain. In Fig.~(\ref{strn37}) we show two examples for one O atom. On the left, we elongated the two Si bonds close to the O atom of 3.7\% with respect to the unstrained LE structure. The strained structure is labelled LS$_{+3.7\%}$. On the right, instead we changed two bonds far from the O atom of about -2.6\% (compression) and 8.3\% (elongation). This structure is labelled LS$^{+8.3\%}_{-2.6\%}$.
\item[2)]  Global Strain (GS): we changed the lattice parameters of the LE $\Sigma$3\{111\} Si GB in the x and y directions leaving the z direction unchanged. As the elongation or compression is rigidly applied to the whole structure, the bond lengths are elongated or compressed of about the same amount. Examples are shown in Fig.~(\ref{cell_strain}) where an elongation of +2.0\% (left) and a compression of -2.0\% (right) were applied to the $a$ and $b$ lattice parameters of the LE structure. These structures are labelled respectively GS$_{+2.0\%}$ and GS$_{-2.0\%}$ and the bond lengths are respectively elongated by about +2.0\% and compressed by about -2.0\%.
\end{itemize}

Applying LS and GS methodologies to the LE $\Sigma$3\{111\} Si GB, we realised a series of strained structures which differs by the percentage of strain and for which we calculated the segregation energy of the O atoms as :

\begin{equation}
\Delta^{n\text{OSGB}}_{n\text{OGB}}  = E^{n\text{OSGB}} - E^{n\text{OGB}},
\label{bulkseg} 
\end{equation}

where $E^{n\text{OSGB}}$ is the impurity energy of the O atoms in the strained GB calculated as :

\begin{equation}
E^{n\text{OSGB}} = E_{n\text{O+SGB}}- E_{\text{SGB}}  - n \mu_{\text{O}},
\label{bulkgb} 
\end{equation}

and $E^{n\text{OGB}}$ is defined in Eq.~(\ref{EnOGB}). In particular, $E_{n\text{O+SGB}}$ is the energy of the strained GB containing $n$ number of O atoms and $E_{\text{SGB}}$ is the energy of the strained GB. 

\begin{figure}[h!]
\centering
\includegraphics[scale=0.4]{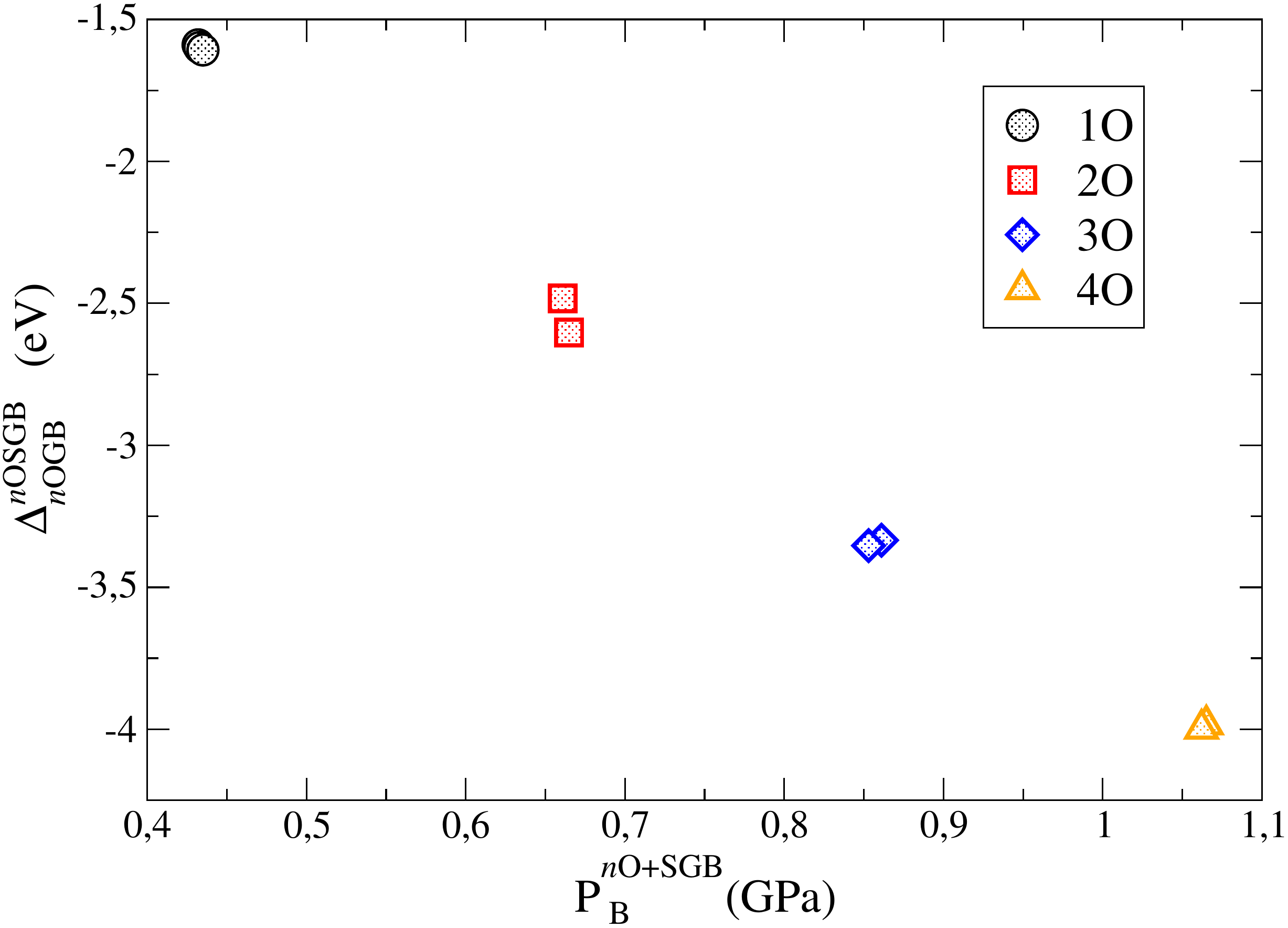}
\caption{Oxygen segregation energy $\Delta^{n\text{OSGB}}_{n\text{OGB}}$ in locally strained GB as a function of P$^{n\text{O+SGB}}_B$. The number of O atoms is $n$ = 1, 2, 3 and 4.}
%Strain values on the bond length in percentage are reported in Table~ \ref{DeltanO}. The different configurations are classified by their pressure (P$_{n\text{O+SGB}}$) with respect to Si bulk (P$_{\text{B}}$)}.}
\label{seg_pressure_plotLS}
\end{figure}

\begin{figure}[h!]
\centering
\includegraphics[scale=0.4]{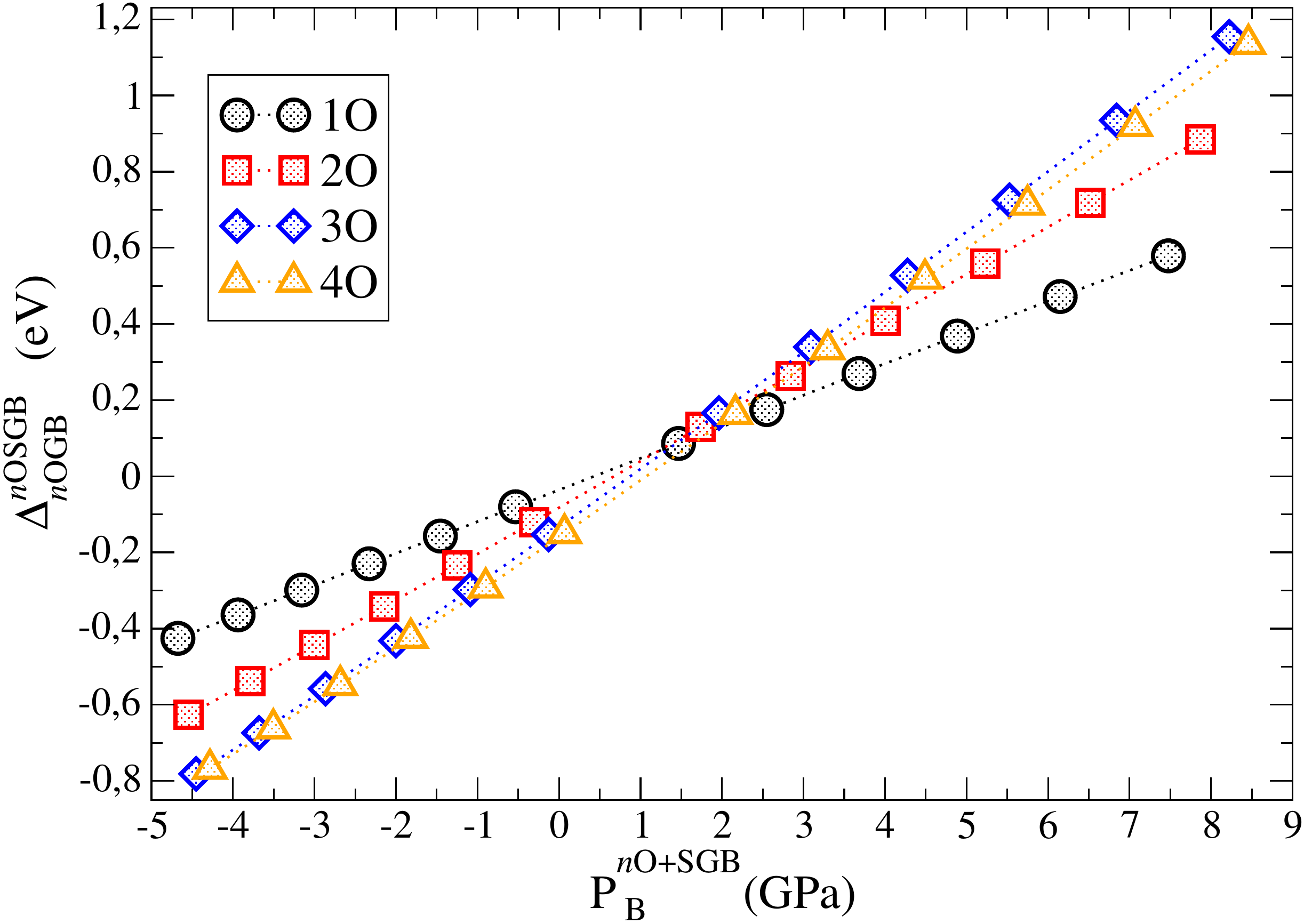}
\caption{Oxygen segregation energy $\Delta^{n\text{OSGB}}_{n\text{OGB}}$ in globally strained GB as a function of P$^{n\text{O+SGB}}_B$. The number of O atoms is $n$ = 1, 2, 3 and 4.}
%Calculated segregation energy of interstitial O atoms  in $\Sigma$3\{111\} GB under global strain ($\Delta^{n\text{OSGB}}_{n\text{OGB}}$). $n$ corresponds to the number of interstitial O atoms. Strain values on cell parameters in percentage are reported in Table~ \ref{DeltanO}. The different configurations are classified by their pressure (P$_{n\text{O+SGB}}$) with respect to Si bulk (P$_{\text{B}}$). Dotted lines are guide for eyes. }}
\label{seg_pressure_plotGS}
\end{figure}

To keep the effect of the strain during the simulation, it is important not to fully relax the strained structures, otherwise the systems would relax to their unstrained counterpart. \cite{PhysRevB.92.075204} Therefore considering first the strained GB with only one O atom we tried two different strategies to calculate the strained total energies : 1) performing a self-consistent calculation, 2) optimising the structure by constraining the Si atoms around the O atom. The results obtained for $\Delta^{\text{1OSGB}}_{\text{1OGB}}$ with the two strategies are reported in the Supp. Mat. and differ by about $3\times10^{-3}$ eV which is also two orders of magnitude smaller than the DFT accuracy. For these reasons, for $n$ = 2, 3 and 4 O atoms, we proceeded by just performing a self-consistent calculation.

In Fig.~(\ref{seg_pressure_plotLS}) we show the O segregation energies ($\Delta^{n\text{OSGB}}_{n\text{OGB}}$) in LS GBs as a function of their pressure $P^{n\text{O+SGB}}_{\text{B}}=P_{n\text{O+SGB}} - P_{\text{B}}$, calculated as the difference between the pressure of the SGB with $n$ number of O atoms ($P_{n\text{O+SGB}}$) and the pressure of the Si bulk ($P_{\text{B}}$). We found that even if locally we have a tensile strain close to the O atom, the $P_{n\text{O+SGB}}$, which average over the whole cell, is always positive, indicateing therefore  a compressive strain. \cite{PhysRevB.92.075204} It is important to underline this point because we obtain that all the structures have negative $\Delta^{n\text{OSGB}}_{n\text{OGB}}$ which means that O segregation is expected. However, the segregation is strongly related to the local tensile strain around the O atoms.
The local geometry of the boundary site where the O is segregated seems to be clearly correlated to the capacity of the GB to act as a gettering site. Moreover, the higher is the number of oxygen atoms in the strained region, the more effective is the segregation.
The explicit values of $\Delta^{n\text{OSGB}}_{n\text{OGB}}$ in the case of LS are reported in Tab.~(\ref{DeltanOSGBnOGB}) and in the Supp. Mat. for comparison with GS structures. 

We point out that the values of LS considered here are only those where the strain is applied close to the O atoms. In fact, calculating the effects of strain far from the O atoms we observed that it is much less effective in in fostering the segregation process. To obtain segregation energies of the same order of magnitudes as those in Tab.~(\ref{DeltanOSGBnOGB}) we need to apply much stronger strain. For example for 
LS$^{+8.3\%}_{-2.6\%}$ we obtain $\Delta^{1\text{OSGB}}_{1\text{OGB}} = -1.552$ eV with  $P^{n\text{O+SGB}}_{\text{B}} = 0.466$ GPa.

In Fig.~(\ref{seg_pressure_plotGS}) we show $\Delta^{n\text{OSGB}}_{n\text{OGB}}$ in GS GBs as a function of the pressure $P^{n\text{O+SGB}}_{\text{B}}$. The trend we observe is independent on the number of O atoms. The more the pressure is negative (tensile), the more the O atoms are segregated.  Instead, exactly the opposite is obtained in the case of compression. These effects seem to be progressively more pronounced as the number of oxygen atoms increases and it seems to stabilise for 3 and 4 O atoms. 
This behaviour, including the reversal in the oxygen segregation capacity between tensile and compressive strain, is probably due to the fact that under tensile strain silicon and oxygen atoms can mimick the structure of SiO$_2$ which in fact has a lattice parameter larger than Si: this cannot happen in the case of compression.
The explicit values of $\Delta^{n\text{OSGB}}_{n\text{OGB}}$ are reported for some of the structures in Tab.~(\ref{DeltanOSGBnOGB}), while in the Supp. Mat. we reported all the values.

We observe that for the GS, contrary to the LS, there is no ambiguity in the interpretation of the sign of the pressure: an elongation (compression) of the lattice parameters implies an elongation (compression) of the bonds. 

\begin{table}[h!]
\begin{center}
\begin{tabular}{ |c|c|c||c|c||c|c||c|c|c| } 
\hline
 & $P^{\text{1O+SGB}}_{\text{B}}$ & $\Delta^{\text{1OSGB}}_{\text{1OGB}}$ & $P^{\text{2O+SGB}}_{\text{B}}$ & $\Delta^{\text{2OSGB}}_{\text{2OGB}}$ & $P^{\text{3O+SGB}}_{\text{B}}$ &  $\Delta^{\text{3OSGB}}_{\text{3OGB}}$ & $P^{\text{4O+SGB}}_{\text{B}}$&  $\Delta^{\text{4OSGB}}_{\text{4OGB}}$  \\ 
\hline
LS$_{+3.7\%}$ & 0.432 & -1.589 & 0.665 & -2.602& 0.861 &-3.334 &  1.065  & -3.984\\ 
LS$_{+5.7\%}$ & 0.435 & -1.607 & 0.661 & -2.482 & 0.853 &-3.353 &  1.062 & -3.999 \\ 
GS$_{+3.0\%}$ & -4.676 & -0.426 & -4.543 &  -0.626 & -4.452 & -0.781 &-4.282& -0.768\\ 
GS$_{+0.5\%}$ & -0.533 & -0.080 &  -0.308 & -0.120 & -0.126   & -0.153 &-0.067 & -0.150  \\ 
GS$_{-0.5\%}$ & 1.467 & 0.086 & 1.735 & 0.129 &1.962  & 0.166 &  2.163& 0.163  \\ 
GS$_{-3.0\%}$ & 7.476 & 0.579 & 7.870 & 0.884 &  8.223   & 1.154 & 8.458 & 1.135\\ 
\hline
\end{tabular}
\end{center}
\caption{Oxygen segregation energy  $\Delta^{n\text{OSGB}}_{n\text{OGB}}$ (eV) in locally and globally strained GBs and $P^{n\text{O+SGB}}_{\text{B}}$ (GPa) for $n$ equal to 1,2,3 and 4 oxygen atoms. }
\label{DeltanOSGBnOGB}
\end{table}

From Tab.~(\ref{DeltanOSGBnOGB}) it's clear that oxygen segregation is strongly disadvantaged in the case of  compressive strain (GS$_{-0.5\%}$, GS$_{-3.0\%}$), while it is favorite in the case of tensile strain for both LS and GS and with increasing number of O atoms. \cite{KasinnoJAP13,OhnoAPL2017} Basically, the homogeneous/inhomogeneous distribution of tensile strain field around segregating sites is playing a role both in case of LS and GS, where the effect is more prominent in the case of the former than the latter.
In the case of LS, we have the maximum inhomogeneous distribution of stress, whereas in case of GS the applied stress distributes homogeneously over all sites, hence its relative distortion is weak. Now as we increase the numbers of oxygen atoms, the local distortion is even more pronounced if compared to isolated impurity and hence serving effectively on the process of segregation. 

In Fig.~(\ref{All_DOS_he_le-crop}) we plotted the DOS of the strained structures to analyse how the electronic properties change with respect to the pure unstrained GB and unstrained LE structures. We observe that the electronic properties are almost unchanged despite the fact that important segregation energies have been calculated in some of the strained GBs. This can be explained by the fact that silicon atoms in all the structures continue to keep their tetrahedral coordination. Only the GS induces some changes in the structure, which however seem to remain quite small.

\subsection{Oxygen atoms in $\Sigma$3\{111\} Si GB with a Si vacancy}
\label{OinterstitalsVacancy}

Oxygen atoms can also segregate in Si GBs in the presence of Si vacancies. In fact, when a vacancy is created in the GB, it induces coordination defects in the structure, which means that Si is not four-fold coordinated anymore. Therefore, Si vacancies create dangling bonds which modify both the energetic and the electronic properties of the system. 

\begin{figure}[h!]
\centering
\includegraphics[scale=0.5,angle=90]{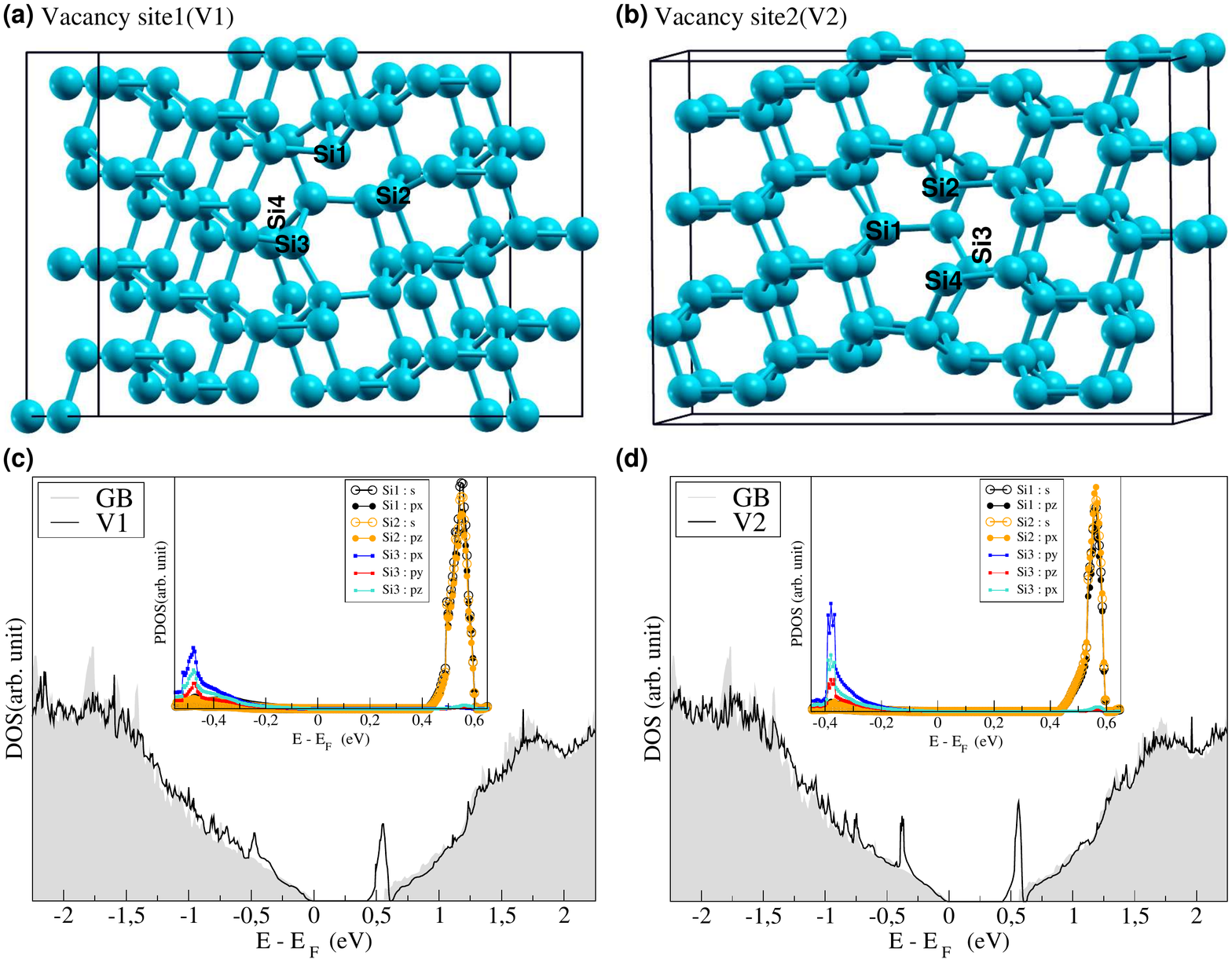}
\caption{Relaxed $\Sigma$3\{111\} Si GB with silicon vacancies labelled as $V1$ (a) and $V2$ (b). All Si atoms are four-fold coordinated except the atoms we labelled as Si1 and Si2 which are three-fold coordinated. On the bottom the DOS of the corresponding above structures ((c) $V1$, (d) $V2$) are reported with a focus in the insets that shows the projected DOS (P-DOS) of Si1, Si2 and Si3 atoms (P-DOS of atom Si4 is not reported because it is superimposable to that of atom Si3.}
\label{Single_vacancy}
\end{figure}

Within $\Sigma$3\{111\} Si GB only two distinct positions of Si exist, where vacancies can be created. These vacancies are shown in Fig.~(\ref{Single_vacancy}) and indicated throughout the paper as $V1$ and $V2$. The effect of creating a vacancy is that, after structural relaxation, two Si atoms become three-fold coordinated. These atoms are labelled as Si1 and Si2.

The formation energy of the vacancy in the GB is calculated as
\begin{equation}
E^{\text{f}}_{\text{VGB}} = E_{\text{VGB}} - \frac{n_{\text{GB}}-1}{n_{\text{GB}}}  E_{\text{GB}}, 
\label{deltanXVGB} 
\end{equation}
where $E_{\text{VGB}}$ is the total energy of the GB with a vacancy, $n_{\text{GB}}$ and $E_{\text{GB}}$ are defined in Eq.~(\ref{EfGB}).\cite{KasinnoJAP13} For $V1$ vacancy we found $E^{\text{f}}_{\text{VGB}} = 2.91$ eV, while for $V2$ vacancy we found $E^{\text{f}}_{\text{VGB}} = 3.04$ eV.

In Fig.~(\ref{Single_vacancy}) the DOS of $V1$ and $V2$ is also shown. Both $V1$ and $V2$ present a sharp peak at the bottom of the conduction which by the analysis of the projected-DOS (see inset) is due to the three-fold coordinated atoms Si1 and Si2. Another peak is also observed at around -0.5 eV in the valence band. This peak comes from the atoms Si3 and Si4 indicated on the so called $V1$ and $V2$ structures. Si3 and Si4 are four coordinated atoms but the presence of the vacancy in the GB induces strong structural distortions that are at the origin of this peak. In the pristine GB the atoms Si3 and Si4 are four-fold coordinated with angles $\sim$ 109.5 $^{\circ}$ and bond lengths $\sim$ 2.35 \AA, while in the presence of the vacancies the angles are $\sim$ 58.0 $^{\circ}$, $\sim$ 122.5 $^{\circ}$, $\sim$ 122.3 $^{\circ}$  and $\sim$ 115.1 $^{\circ}$ and the bond lengths change of $\sim$ 0.2 \AA.

\begin{figure}[h!]
\centering
\includegraphics[scale=0.2]{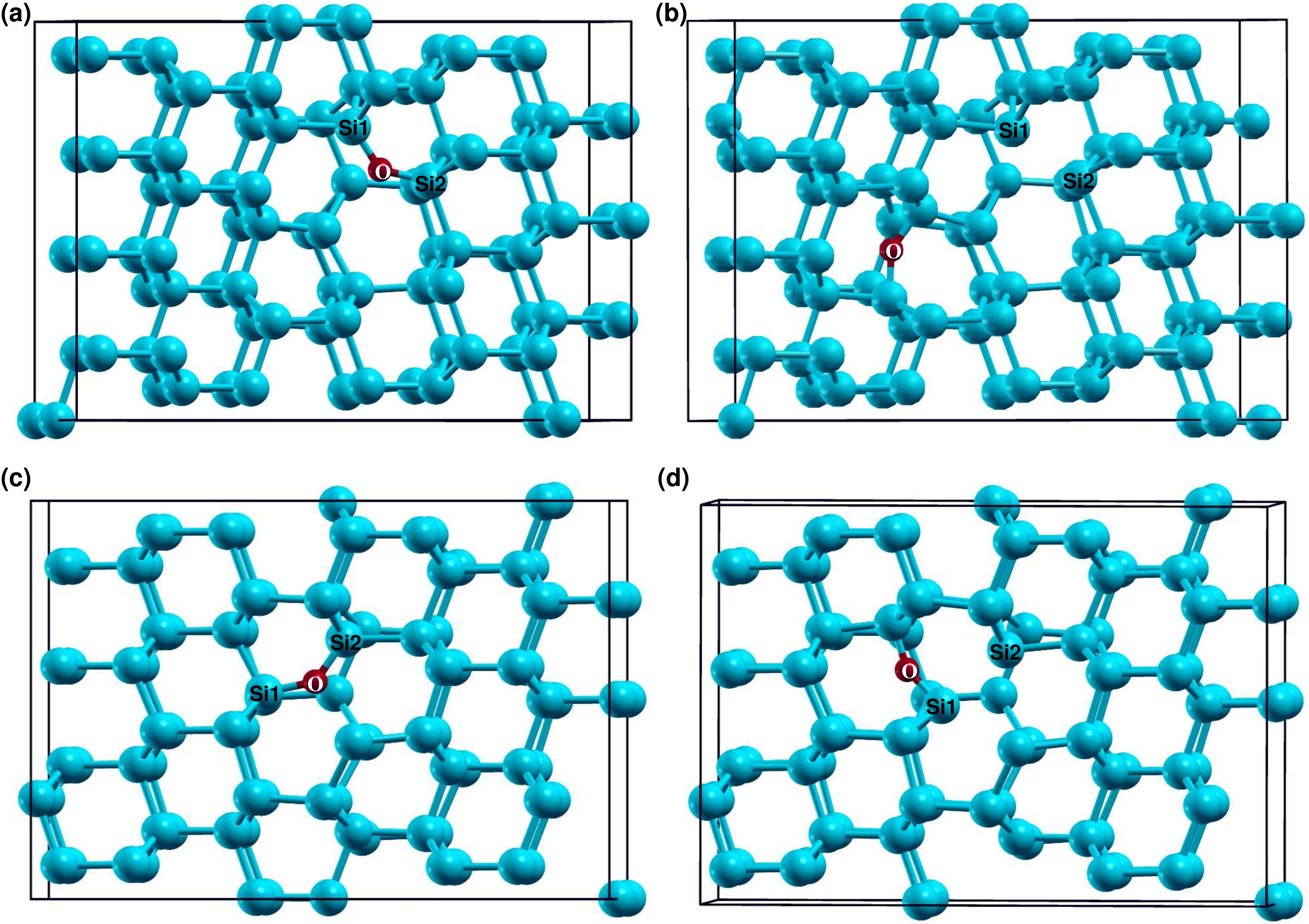}
\caption{$\Sigma$3\{111\} Si GB with $V1$ ((a) and (b)) or $V2$ ((c) and (d)) vacancy and one O atom (red ball). In a) and c) the O atom creates bonds with the three-fold coordinated Si atoms (Si1 and Si2) restoring the four coordination for all the atoms. In (b) and (d), the O atom places at bond-centered position between two four-fold coordinated Si atoms. The atoms Si1 and Si2 are still three-fold coordinated.}
\label{SiVgb_1O}
\end{figure}

Starting from $V1$ and $V2$ vacancies, we inserted O atoms in different configurations and we optimised the structures. We found that two opposite situations can occur that we show in Fig. (\ref{SiVgb_1O}) for one O atom. We obtained that: 1) the O atom  bonded with the three-fold coordinated Si1 and S2 atoms restoring the four coordination (no dangling bonds), 2) the O atom placed at bond-centered position between two four-fold coordinated Si atoms (two dangling bonds). In the case of two, three and four O atoms, we obtained the same two situations described for a single oxygen with the rest of the oxygen atoms in bond-centered positions. 

The segregation energy of the O atoms at the GB with a vacancy is
\begin{equation}
\Delta^{n\text{OVGB}}_{n\text{OGB}} = E^{n\text{OVGB}} - E^{n\text{OGB}} 
\label{deltanXVGB} 
\end{equation}
where $E^{n\text{OVGB}}$ and $E^{n\text{OGB}}$ are the impurity energies of the O atoms respectively in the GB with a vacancy and in the GB which are calculated as
\begin{equation}
E^{n\text{OVGB}} = E_{n\text{O+VGB}}- E_{\text{VGB}}  - n \mu_{\text{O}}, 
\label{EnOVGB} 
\end{equation}
where $E_{n\text{O+VGB}}$ is the energy of the GB with a vacancy with $n$ atoms of O, $E_{\text{VGB}}$ is the energy of the GB with a vacancy (VGB), $\mu_{\text{O}}$ is the chemical potential of the O atom, and $E^{n\text{OGB}}$ is defined in Eq.~\ref{EnOGB} (Sec.~\ref{OinterstitalsStrain}). 

\begin{figure}[h!]
\centering
\includegraphics[scale=0.4]{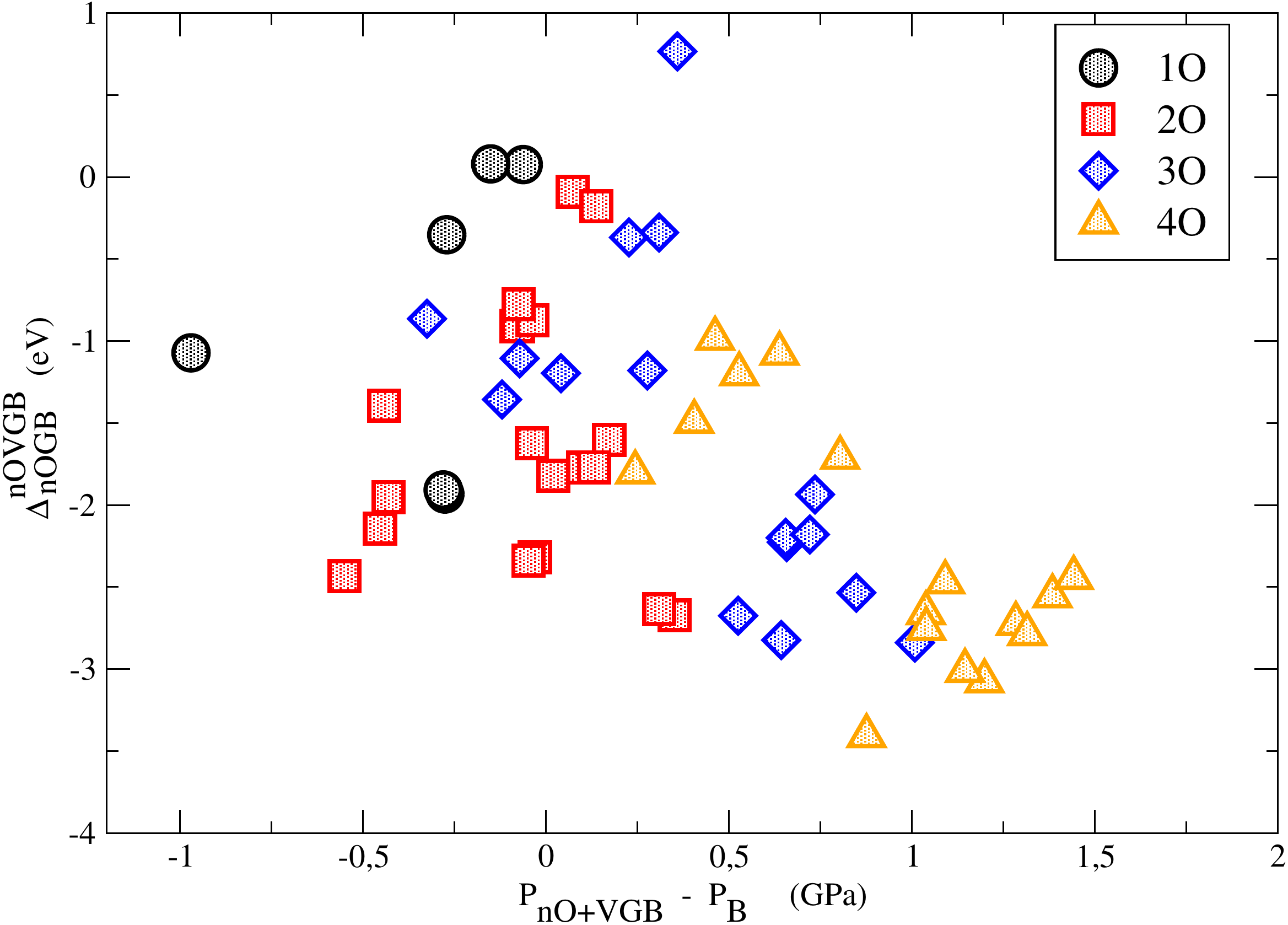}
\caption{O segregation energy $\Delta^{n\text{OVGB}}_{n\text{OGB}}$ as a function of P$^{n\text{O+VGB}}_\text{B}$. The number of O atoms is $n$ = 1, 2, 3 and 4.}
\label{seg_pressure_plot_V}
\end{figure}

In Fig.~(\ref{seg_pressure_plot_V}) we show the O segregation energies $\Delta^{n\text{OVGB}}_{n\text{OGB}}$ for the investigated structures as a function of their pressure $P^{n\text{O+VGB}}_{\text{B}}=P_{n\text{O+VGB}} - P_{\text{B}}$ calculated as the difference between the pressure of the GB with the vacancy containing $n$ number of O atoms ($P_{n\text{O+VGB}}$) and the pressure of the Si bulk ($P_{\text{B}}$) . The explicit values of $\Delta^{n\text{OVGB}}_{n\text{OGB}}$ are reported in Tab.~(\ref{nOVGBnOGB}) for the structures with the highest and lowest segregation energy. The values for all the structures are reported in the Supp. Mat.. Similarly to strained structures, the segregation energy progressively decreases including one, two, three and four O atoms.

\begin{table}[h!]
\begin{center}
\begin{tabular}{ |c|c|c|c|c|c|c|c|c| } 
\hline
&$P^{\text{1O+VGB}}_{\text{B}}$ & $\Delta^{\text{1OVGB}}_{\text{1OGB}}$ &$P^{\text{2O+VGB}}_{\text{B}}$ & $\Delta^{\text{2OVGB}}_{\text{2OGB}}$& $P^{\text{3O+VGB}}_{\text{B}}$ &  $\Delta^{\text{3OVGB}}_{\text{3OGB}}$&$P^{\text{4O+VGB}}_{\text{B}}$ &  $\Delta^{\text{4OVGB}}_{\text{4OGB}}$ \\ 
\hline
\hline
HE$_{V1}$  &-0.062 & 0.075&0.073 &-0.091 & 0.359  &0.766 &0.462 & -0.985 \\ 
LE$_{V1}$  &-0.276 & -1.933 & 0.352& -2.671 &1.008 &-2.839 & 1.198  & -3.073\\ 
 HE$_{V2}$ & -0.151 & 0.080 & 0.138 &-0.179 & 0.325 & -0.865  &0.638  & -1.076  \\ 
LE$_{V2}$  &-0.281 & -1.908 &0.309 &-2.634 & 0.643  & -2.824  &0.876  & -3.408  \\ 
\hline
\end{tabular}
\end{center}
\caption{Oxygen segregation energy $\Delta^{n\text{OVGB}}_{n\text{OGB}}$ (eV) and $P^{n\text{O+VGB}}_{\text{B}}$ (GPa) for $n$ equal to 1,2,3 and 4 oxygen atoms. The values are reported for the LE (lowest total energy) and HE (highest total energy) for $V1$ and $V2$ vacancies.}
\label{nOVGBnOGB}
\end{table}

\begin{figure}[h!]
\centering
\includegraphics[scale=0.4]{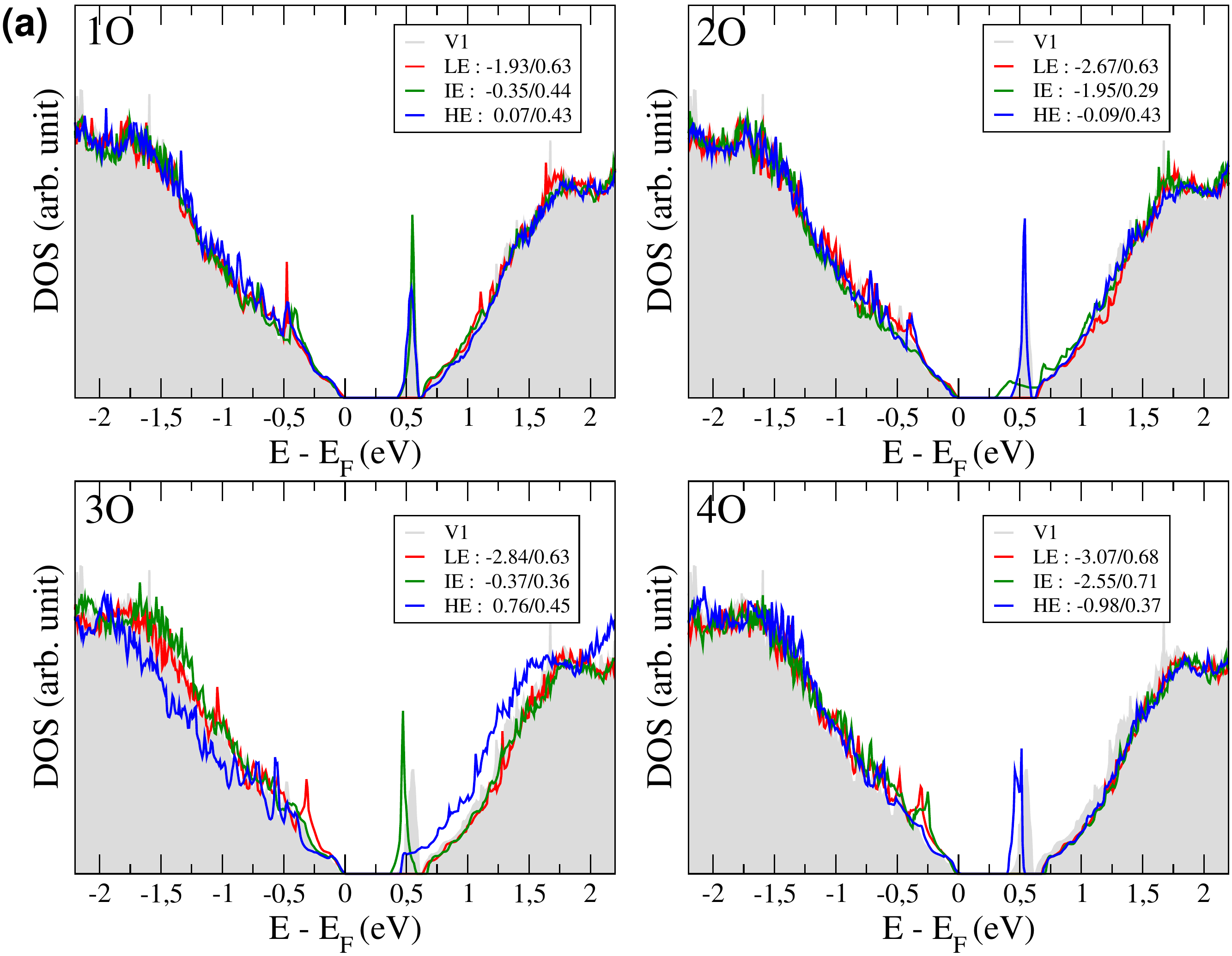}
\includegraphics[scale=0.4]{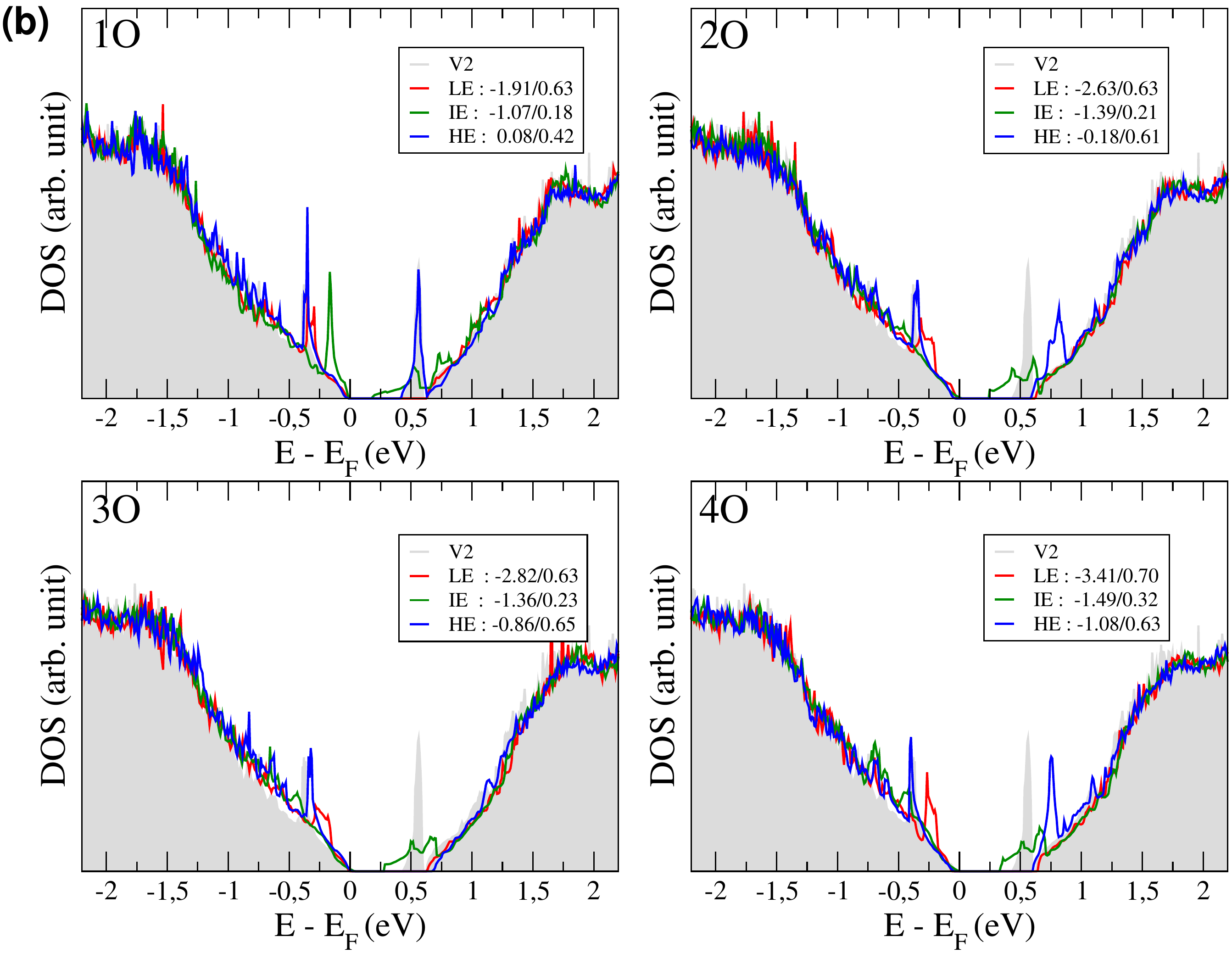}
\caption{The DOS of $V1$ (a) and $V2$ (b) with 1O, 2O, 3O and 4O atoms, respectively. The DOS of the GB with the vacancy only (grey shadow) is compared with that of the same structure plus an increasing number of oxygen atoms in the case of the lowest (LE red line), highest (HE blue line) and intermediate (IE green line) total energy. Numerical values in the inset are the segregation energy and the energy gap in eV, respectively.} 
%(HE in blue) energy configurations of the GB with a Si vacancy and an increasing number of O atoms (from 1 up to 4) compared with the DOS of GB with a Si vacancy (grey shadow)}
\label{Dos_vacancy}
\end{figure}

In Fig.~(\ref{Dos_vacancy}) we plotted the DOS of $V1$ with 1O, 2O, 3O and 4O (top panels) and the DOS of $V2$ with 1O, 2O, 3O and 4O (bottom panels). We selected three configurations: the structures with the lowest total energy (LE), with the highest total energy (HE) and with intermediate total energy (IE) between LE and HE.

The behaviour of the DOS can be interpreted as following. All the LE structures have the dangling bonds saturated by an O atom (no dangling bonds). The peak at the bottom of the conduction band is therefore not present anymore. Instead, the peak in the valence band due to geometrical distortion of four-fold Si atoms in the presence of the vacancy is still present. These types of structures are the most energetically stable and have the electronic gap similar to the GB electronic gap.

The IE and HE structures have a more complex and different geometrical rearrangement of the Si and the O atoms. At the bottom of the conduction band we can find or not the peak already present in the pure $V1$ and $V2$ structures. The existence of this peak is always related to the fact that the O can saturate or not the dangling bonds of the Si atoms. Moreover, the peak, if present, can be sharp or broad. This is due to modification of the geometric structure such as bond lengths and angles with respect to their pristine counterpart (the GB with the vacancy only).  The peak in the valence band instead is always due to the four-fold silicon atoms and relative structural distortions. The IE and HE structures can strongly decrease the electronic gap as many different probable segregation sites investigated have shown.

\subsection{Discussion}	

The segregation energy mechanisms depend on local geometry distortions obtained by the GBs itself or otherwise generated by strain and/or vacancies.  In $\Sigma$3\{111\} Si GBs, we found that O atoms do not segregate as well as other kinds of impurities such as C, B, P and Co, even if the impurity concentrations were high.  \cite{ohnoapl2013,OhnoAPL2016} Instead, less symmetric boundaries such as $\Sigma$9\{221\} and $\Sigma$27\{552\} demonstrated to be efficient gettering centers for atomic impurities. \cite{ZHAO2017599,ZHAO201952,PhysRevLett.121.015702,PhysRevLett.77.1306,OhnoAPL2016} It is therefore clear that segregation is expected when the system is locally distorted. 

The strain is present in GBs and induces additional local distortions. For oxygen atoms, experimental studies predict O atoms to segregate under tensile stress to attain more stable bonding network by reducing the local stress favourably in higher order grain boundary. \cite{OhnoAPL2017,OhnoAPL2016} Through our calculations, we confirm that for O, tensile strain is always favorite and that local strain seems to be more effective than global strain. The role of strain was also pointed out for other atomic impurities such as C, B, Al, Ge, Fe and P. \cite{PhysRevB.91.035309,YuAndreyJAP2015,KasinnoJAP13,SinnoJAP2015,ZHAO201952} However, it was found that for some impurities it is the compressive strain that favours segregation. 

The presence of vacancies also induces local geometrical deformation of the GB. \cite{Feng2009,PhysRevB.37.7482,PhysRevB.58.1318,ZHANG2020144437,YuAndreyJAP2015} The segregation is favorite for those structures which have restored tetrahedral covalent bonds. The impurities become substitutional defects. \cite{PhysRevB.91.035309} The presence of vacancies attract atomic impurities in order to restore the electronic stability. The same behaviour was also observed in Ni GBs with Si as impurities. \cite{Mazalova2020}
 
\begin{figure}[h!]
\centering
\includegraphics[scale=0.4]{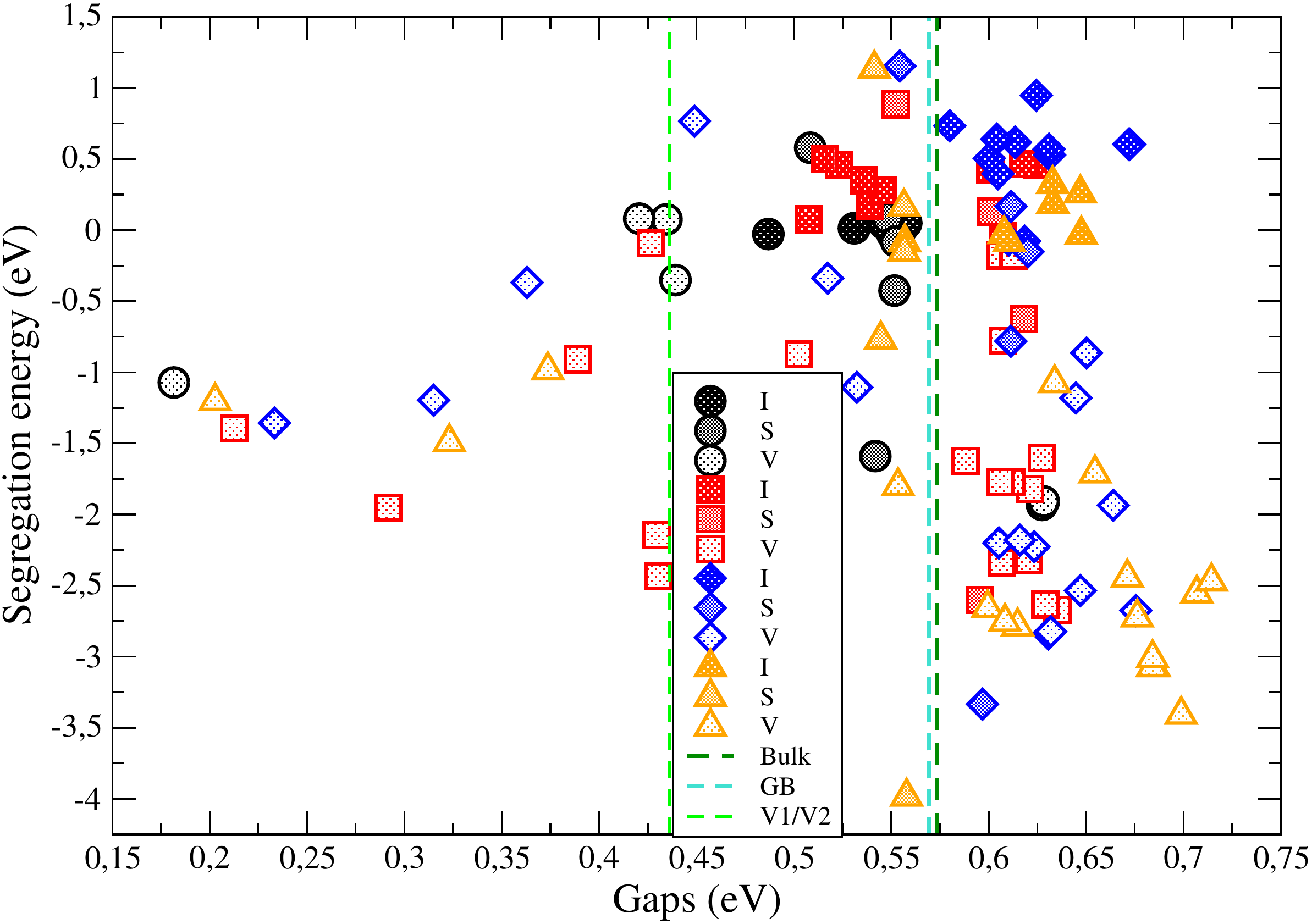}
\caption{The segregation energies for the $\Sigma$3\{111\} Si GB in presence of interstitial impurities (I, solid symbols), strain (S, heavy shading) and vacancies (V, light shading) as a function of the relative electronic band gaps. As a reference we plotted with dashed lines the Si bulk (dark green), the $\Sigma$3\{111\} Si GB (cyan) and the V1 and V2 (light green) electronic band gaps. Structures with increasing number of oxygen atoms are represented with different colours and symbols: 1O black circles, 2O red squares, 3O blue diamonds, 4O orange triangles. }
%(HE in blue) energy configurations of the GB with a Si vacancy and an increasing number of O atoms (from 1 up to 4) compared with the DOS of GB with a Si vacancy (grey shadow)}
\label{Gaps}
\end{figure}

The electronic properties of the $\Sigma$3\{111\} Si GB are not significantly changed for interstial oxygen atoms and in the presence of strain. The DOS has the same shape as for the pure GB. The oxygen atoms segregate at bond-centered position between two silicon atoms, restoring the configuration of SiO$_2$. The electronic properties are therefore determined by the $\Sigma$3\{111\} Si GB.

Instead, a very different situation occurred in the presence of vacancies. A vacancy creates dangling bonds which can interact with the oxygen atoms segregating at the GB. The way vacancies interact with atomic impurities can be variegated and strongly depends on the number of segregated species. Predicting and then tuning the electronic behaviour is very complicated. In fact, there exist cases where oxygen atoms can passivate dangling bonds coming from vacancies, suppressing electron-hole recombination and cases where instead deep energy states still exist.
 
In Fig.~(\ref{Gaps}) we show the segregation energies for the $\Sigma$3\{111\} Si GB in presence of interstitial impurities, strain and vacancies as a function of the electronic band gaps. As a reference we also plotted the Si bulk, the 
$\Sigma$3\{111\} Si GB and the V1 and V2 electronic band gaps (V1 and V2 gaps are almost the same). The gaps of  strained structures or with the presence of interstitial impurities change very little with respect to Si bulk and $\Sigma$3\{111\} Si GB (approximatively of about $\pm$ 0.05 eV). The behaviour is very different in the case of vacancies: actually the energy gaps of pure V1 and V2 structures is significantly smaller than that of bulk Si or GB. Moreover the presence of interstitial impurities coupled with the vacancy spreads the energy gaps in a wide range of about $\pm 0.3$ eV around the reference value.
In terms of segregation energies it can be observed that while oxygen impurities are not able to segregate in pristine $\Sigma$3\{111\} Si GB the presence of strain can, in some cases, favours the segregation while oxygen is always able to segregate  in the presence of a Si vacancy.

%\blue{Zhao 52e62.pdf  there are no energy states in the bandgap, being in accordance with the absence of dangling bonds in these GBs. one can find that deep states exist in the bandgap.
%Further calculations of the partial den- sity of states (PDOS) at different core sites at the GB reveal that only Si atoms at site 5 generate these strong deep states (cf. Fig. 6b), contributing to the most part of the deep levels in the bandgap in S-  GB. 
%Deep levels in the bandgap which result in significant P and As segregation, can be created not only by the extra bonds, but also by dangling bonds of the atomic sites at the Si GBs. deep %levels in XXX GBs have a mixed nature, which are produced both by the dangling bonds of sites 1, and by the extra bonds of sites 6 (cf. Fig. 7a). However, the deep levels in XXX These %deep levels created either by extra bonds or dangling bonds can trap or localize electrons and make these GB sites exhibit a high attractive binding energy for n-type dopants. As a result %of the deep levels, substantial P and As segregations at the Se and XXX GBs can be expected. 
%}
 
\section{Conclusions}

We studied from first-principles the role of strain and vacancies in the segregation energy of O atoms at $\Sigma$3\{111\} Si GB. We considered tensile and compressive strain locally and globally applied to the GB. Then, multiple oxygen atoms were introduced in the strained GBs in many different possible configurations. For each configuration and type of strain, we calculated the oxygen segregation energy revealing the factors influencing the segregation itself. We found that oxygen segregation is always favoured by tensile strain and that local strain seems to be more effective for segregation than global one. The same methodology was then implemented to investigate the oxygen segregation in $\Sigma$3\{111\} Si GB with a Si vacancy: we included multiple oxygen atoms in different structural configurations, trying to deduce the mechanisms that regulate the process. The segregation is favorite for those structures which have restored a full tetrahedral coordination. The interstitial oxygen atoms, attracted by the presence of vacancies, become substitutional defects that tend to restore the electronic stability.  The inhomogeneous distribution of strain field around segregating sites, due to both local geometrical distortion or the presence of a Si vacancy, is the main factor that influences the oxygen segregation at GBs.

Then we studied how the electronic properties of GBs can change due to O segregation. The correlation between strain and vacancies and the electronic structure of the GBs was pointed out. For each structure we have analysed the DOS and the band gaps versus the segregation energy as well as the projected DOS in order to characterise the origin of new energy levels. Actually, the knowledge of the nature of particular electronic states enables researchers to design new materials and strategies to reduce non radiative electron-hole recombination avoiding charge and energy losses and therefore improving solar cell efficiency.

A full understanding of the role of GBs and impurities in the solar cells performance is a very difficult problem for which further theoretical and experimental investigations would be required. The role of different types of GBs (atoms/size/order) and impurities on the electronic properties will be a future step in order to calculate and optimize the optical absorption of devices.

\section{Acknowledgements}
We would like to thank the University of Modena and Reggio Emilia for the financial support (FAR 2018), as well as the CINECA HPC facility for the approved ISCRA C project ESGB-Si (IsC79-ESGB-Si).

\bibliography{bib}

\end{document}

\title{Supplementary Material : \\
Ab initio study of oxygen segregation in silicon grain boundaries: the role of strain and vacancies}
\author{Rita Maji}
\affiliation{Dipartimento di Scienze e Metodi dell'Ingegneria, Universit{\`a} di Modena e Reggio Emilia,Via Amendola 2 Padiglione Morselli, I-42122 Reggio Emilia, Italy}
\author{Eleonora Luppi}
\affiliation{Sorbonne Universit\'e, UPMC Univ Paris 06, UMR 7616, Laboratoire de Chimie Th\'eorique, F-75005 Paris, France. CNRS, UMR 7616, Laboratoire de Chimie Th\'eorique, F-75005 Paris, France.}
\author{Nathalie Capron} 
\affiliation{Sorbonne Universit\'e, CNRS, Laboratoire de Chimie Physique Mati\`ere et Rayonnement, UMR 7614, F-75005 Paris, France.}
\author{Elena Degoli}
\affiliation{Dipartimento di Scienze e Metodi dell'Ingegneria, Universit{\`a} di Modena e Reggio Emilia, Via Amendola 2 Padiglione Morselli, I-42122 Reggio Emilia, Italy, \\ 
Centro Interdipartimentale En$\&$Tech, Via Amendola 2 Padiglione Morselli, I-42122 Reggio Emilia, Italy, \\
Centro S3, Istituto Nanoscienze-Consiglio Nazionale delle Ricerche (CNR-NANO),Via Campi 213/A, 41125 Modena, Italy}
%\date{\today}
%\pacs{?}

\maketitle
\setstretch{1}
%\section{Oxygen interstitials in $\Sigma$3\{111\} GB}
%\label{sec:XXX}

\begin{figure}[h!]
\centering
\includegraphics[scale=0.28,angle=0]{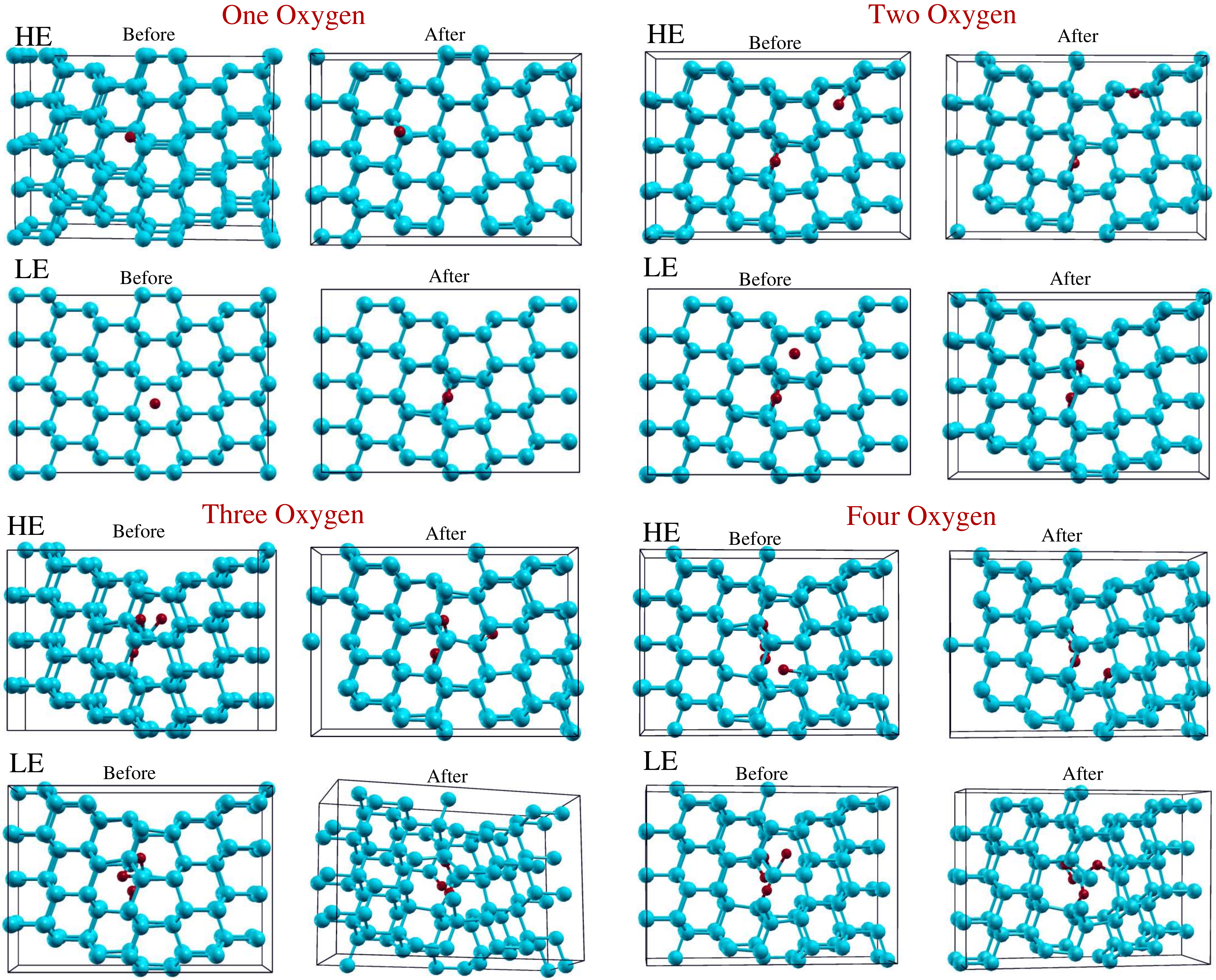}
\caption{HE and LE with oxygen interstitials in $\Sigma$3\{111\} GB before and after optimisation.}
\label{GB96}
\end{figure}

\begin{table}[h!]
\begin{center}
\begin{tabular}{ |c|c|c|c|c|c|c|c|c|c| } 
\hline
 $P^{\text{1O+GB}}_{\text{B}}$ & $\Delta^{\text{1OGB}}_{\text{1OB}}$ & $P^{\text{2O+GB}}_{\text{B}}$ &  $\Delta^{\text{2OGB}}_{\text{2OB}}$ & $P^{\text{3O+GB}}_{\text{B}}$ &  $\Delta^{\text{3OGB}}_{\text{3OB}}$ & $P^{\text{4O+GB}}_{\text{B}}$ & $\Delta^{\text{4OGB}}_{\text{4OB}}$   \\ 
\hline
\hline
 0.363   &  0.016  & 0.687  & -0.039 &  0.891 & -0.078 & 1.091 &  -0.085\\ 
 0.379  &  0.043  & 0.800  &   0.076 & 0.941 & -0.070 & 1.302  &  -0.031 \\
 0.382  &  0.049  & 0.883  &   0.464 & 1.044 &  0.528 &  1.342  &  -0.083    \\
 0.440  & -0.030  & 0.889  &   0.462 & 1.091 &  0.947 & 1.346 &  0.325 \\
 0.499  &  0.013  & 0.937  &   0.425 & 1.107  &  0.396 & 1.352  &  -0.028 \\
 0.500  &  0.013  & 0.952  &   0.279 & 1.123 &  0.619 & 1.404  &  0.259\\ 
 0.511  &  -0.027 & 0.956  &   0.169 & 1.146 &  0.569 && \\
             &               & 0.971  &   0.351 & 1.180 &  0.640 &&\\
             &               & 0.994  &   0.499 & 1.192 &  0.604 &&\\
             &               &  0.997 &   0.456 & 1.197 &  0.614 &&\\
             &               &             &                & 1.218 &  0.504 &&\\
             &               &             &                & 1.257 &  0.739 &&\\
\hline
\end{tabular}
\end{center}
\caption{Oxygen segregation energy $\Delta^{\text{nOGB}}_{\text{nOB}}$ (eV) and $P^{\text{nO+GB}}_{\text{B}}$ (GPa) for $n$ equal to 1,2,3 and 4 oxygens. Different rows correspond to different possible configurations: each configuration is characterised by its pressure. }
\end{table}

\begin{table}
\begin{center}
\begin{tabular}{ |c|c|c|c|c| } 
\hline
 & $^{\text{scf}}\Delta^{\text{1OSGB}}_{\text{1OGB}}$  (eV)  & $^{\text{opt}}\Delta^{\text{1OSGB}}_{\text{1OGB}}$ \\ 
\hline
\hline
LS$_{+3.7\%}$ & -1.589 &-1.586 \\ 
%LS$_{+4.7\%}$ & -1.59887544 &-1.59523720 \\ 
LS$_{+5.7\%}$ & -1.607 &-1.603 \\ 
%LS$^{*}_{+x.x\%}$ & -1.54949448 & \\ 
\hline
\end{tabular}
\end{center}
\caption{$^{\text{scf}}\Delta^{\text{1OSGB}}_{\text{1OGB}}$ and $^{\text{opt}}\Delta^{\text{1OSGB}}_{\text{1OGB}}$ are O segregation energy calculated by performing a simple scf calculation to extract $E_{\text{1O+SGB}}$ or optimised by fixing Si around O atoms.  }
\end{table}

\begin{figure}[h!]
\centering
\includegraphics[scale=0.358]{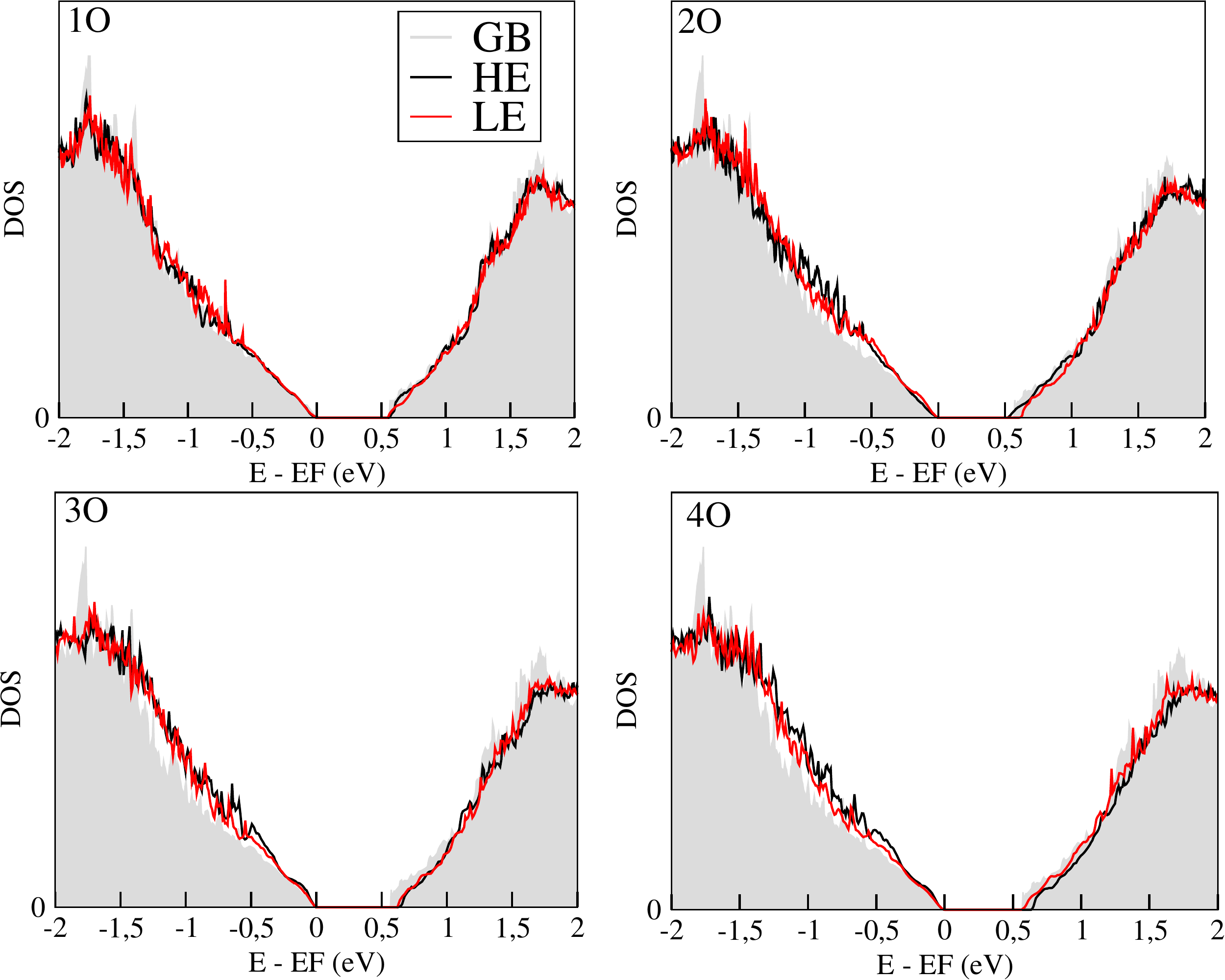}
\caption{Density of states of the $\Sigma$3\{111\} Si GBs without oxygen atoms (grey shadow), or with multiple oxygens in the LE (red) and HE (black) configurations. The number of O atoms is $n$ = 1, 2, 3 and 4.}
\label{All_DOS_he_le-crop}
\end{figure}

\begin{table}[h!]
\begin{center}
\begin{tabular}{ |c|c|c|c|c|c|c|c|c|c| } 
\hline
 & $P^{\text{1O+SGB}}_{\text{B}}$ & $\Delta^{\text{1OSGB}}_{\text{1OGB}}$ & $P^{\text{2O+SGB}}_{\text{B}}$ & $\Delta^{\text{2OSGB}}_{\text{2OGB}}$ & $P^{\text{3O+SGB}}_{\text{B}}$ &  $\Delta^{\text{3OSGB}}_{\text{3OGB}}$ & $P^{\text{4O+SGB}}_{\text{B}}$&  $\Delta^{\text{4OSGB}}_{\text{4OGB}}$  \\ 
\hline
\hline
LS$_{+3.7\%}$ & 0.432 & -1.589 & 0.665 & -2.602 & 0.861 &-3.334 &  1.065  & -3.984\\ 
LS$_{+5.7\%}$ & 0.435 & -1.607 & 0.661 & -2.482 & 0.853 &-3.353 &  1.062 & -3.999 \\ 
GS$_{+3.0\%}$ & -4.676 & -0.426 & -4.543 &  -0.626 & -4.452 & -0.781 &-4.282& -0.768 \\ 
GS$_{+2.5\%}$ & -3.937 & -0.364 &  -3.787 & -0.538 &  -3.680 & -0.674 & -3.505& -0.662  \\ 
GS$_{+2.0\%}$ & -3.156 & -0.299 & -2.998 & -0.443 & -2.864 & -0.558 & -2.684& -0.549 \\ 
GS$_{+1.5\%}$ & -2.330 & -0.230 & -2.143 & -0.342 & -2.001 & -0.432 & -1.818& -0.424 \\ 
GS$_{+1.0\%}$ & -1.456 & -0.157 & -1.250  & -0.235 & -1.089  & -0.297 &  -0.900& -0.292   \\ 
GS$_{+0.5\%}$ & -0.533 & -0.080 &  -0.308 & -0.120& -0.126   & -0.153 &-0.067 & -0.150  \\ 
GS$_{-0.5\%}$ & 1.467 & 0.086 & 1.735 & 0.129 &1.962  & 0.166 &  2.163& 0.163  \\ 
GS$_{-1.0\%}$ & 2.549 & 0.175 & 2.840 & 0.263 &  3.090  & 0.339 & 3.297& 0.333 \\ 
GS$_{-1.5\%}$ & 3.680 & 0.270 & 4.004 & 0.408 &  4.276 & 0.527 &  4.490 & 0.518 \\ 
GS$_{-2.0\%}$ & 4.887 & 0.368 & 5.229 & 0.559 &  5.527 & 0.725 &  5.748& 0.713 \\ 
GS$_{-2.5\%}$ & 6.149 & 0.472 & 6.516 & 0.718 & 6.841 & 0.935 &  7.069& 0.920 \\ 
GS$_{-3.0\%}$ & 7.476 & 0.579 & 7.870 & 0.884 &  8.223   & 1.154 & 8.458 & 1.135\\ 
\hline
\end{tabular}
\end{center}
\caption{Oxygen segregation energy in locally and globally strained GB $\Delta^{\text{nOSGB}}_{\text{nOGB}}$ (eV) and $P^{\text{nO+SGB}}_{\text{B}}$ (GPa) for $n$ equal to 1,2,3 and 4 oxygens. }
\label{DeltanO}
\end{table}

\begin{table}[h!]
\begin{center}
\begin{tabular}{ |c|c|c|c|c|c|c|c|c|c|c| } 
\hline
Vacancy &$P^{\text{1O+VGB}}_{\text{B}}$ & $\Delta^{\text{1OVGB}}_{\text{1OGB}}$ & $P^{\text{2O+VGB}}_{\text{B}}$ & $\Delta^{\text{2OVGB}}_{\text{2OGB}}$ & $P^{\text{3O+VGB}}_{\text{B}}$ &  $\Delta^{\text{3OVGB}}_{\text{3OGB}}$ & $P^{\text{4O+VGB}}_{\text{B}}$&  $\Delta^{\text{4OVGB}}_{\text{4OGB}}$  \\ 
\hline
\hline
& -0.276 & -1.933 & -0.430 & -1.951 & 0.227  &-0.369 &  0.462 & -0.985   \\ 
& -0.271 & -0.352 & -0.081 & -0.909 & 0.309  &-0.338 &  1.198 & -3.073   \\ 
& -0.062 &  0.075 & -0.077 & -0.910 & 0.359  & 0.766 &  1.284 & -2.724   \\ 
&        &        & -0.039 & -1.623 & 0.525  &-2.675 &  1.315 & -2.786   \\
&        &        & -0.037 & -0.872 & 0.658  &-2.226 &  1.384 & -2.555   \\   
V1&      &        & -0.031 & -2.319 & 1.008  &-2.839 &  1.442 & -2.443   \\      
&        &        & -0.030 & -2.319 &        &       &        &\\         	
&        &        &  0.020 & -1.821 &        &       &        &\\ 
&        &        &  0.073 & -0.091 &        &       &        &\\    
&        &        &  0.173 & -1.602 &        &       &        &\\     
&        &        &  0.352 & -2.671 &        &       &        &\\                                                                
\hline
&\red{-0.970} & \red{-1.072} & \red{-0.551} & \red{-2.434} & \red{-0.120} &  \red{-1.356} &  \red{0.244} & \red{-1.800} \\
&\red{-0.281} & \red{-1.908} & \red{-0.455} & \red{-2.144} & \red{-0.072} &  \red{-1.105} & \red{0.405}  & \red{-1.491} \\
&\red{-0.280} & \red{-1.908} & \red{-0.453} & \red{-2.143} & \red{ 0.041} &  \red{-1.196} & \red{0.528}  & \red{-1.201} \\
&\red{-0.151} & \red{ 0.080} & \red{-0.442} & \red{-1.394} & \red{0.277}  &\red{ -1.180}  & \red{0.638}  & \red{-1.076} \\
&       &        & \red{-0.075} & \red{-0.776} & \red{0.325}  & \red{-0.865}  & \red{0.804}  & \red{-1.710} \\
\red{V2}&     &        & \red{-0.048} & \red{-2.339} & \red{0.643}  & \red{-2.824}  & \red{0.876}  & \red{-3.408} \\
&       &        &  \red{0.102} & \red{-1.772} & \red{0.655}  & \red{-2.200}  & \red{1.038}  & \red{-2.659} \\
&       &        & \red{0.131}  & \red{-1.769} & \red{0.721}  & \red{-2.180}  & \red{1.039}  & \red{-2.755} \\
&       &        & \red{0.138}  & \red{-0.179} & \red{0.735}  & \red{ -1.935} & \red{1.091}  & \red{-2.470} \\
&       &        &  \red{0.309} & \red{-2.634} & \red{0.848}  & \red{-2.535}  & \red{1.145}  & \red{-3.010} \\

\hline
\end{tabular}
\end{center}
\caption{Oxygen segregation energy in presence of a Si vacancy $\Delta^{\text{nOVGB}}_{\text{nOGB}}$ (eV) and $P^{\text{nO+VGB}}_{\text{B}}$ (GPa) for $n$ equal to 1,2,3 and 4 oxygens. Different rows correspond to different possible configurations: each configuration is characterised by its pressure. }
\label{DeltanO}
\end{table}